%% file: main.tex
\documentclass{article}

\usepackage{type1cm}        
\usepackage{makeidx}         
\usepackage{graphicx}        
\usepackage{multicol}        
\usepackage[bottom]{footmisc}

\usepackage{newtxtext}       %
\usepackage{newtxmath}       
\usepackage{graphicx}
\usepackage{import}
\usepackage[subpreambles=false]{standalone}
\usepackage{float}
\usepackage{subcaption}
\graphicspath{{../}}

\makeindex             

\begin{document}

\title{Phase space learning with neural networks}

\author{Jaime Lopez Garcia\\jaime.lopez.garcia@repsol.com,Repsol Technology Center
\and Angel Rivero Jimenez\\angel.rivero@repsol.com, Repsol Technology Center}

\maketitle

\abstract{This work proposes an autoencoder neural network as a non-linear generalization of projection-based methods for solving Partial Differential Equations (PDEs). The proposed deep learning architecture presented is capable of generating the dynamics of PDEs by integrating them completely in a very reduced latent space without intermediate reconstructions, to then decode the latent solution back to the original space. The learned latent trajectories are represented and their physical plausibility is analyzed. It is shown the reliability of properly regularized neural networks to learn the global characteristics of a dynamical system's phase space from the sample data of a single path, as well as its ability to predict unseen bifurcations.}

\import{sections/}{Introduction}

\import{sections/}{Related_work}

\import{sections/}{PartI}

\import{sections/}{PartII}

\import{sections/}{Conclusion}

\bibliographystyle{unsrt}
\bibliography{main}

\end{document}

%% file: sections/Introduction.tex
\section{Introduction}
Despite the constant improvement in computing power, the intrinsic degrees of freedom present in some regimes of complex non-linear model solutions, such as turbulent flows, make them unsuitable for control applications, or extremely difficult to solve for complex geometries. Recently, there is ongoing research to find more reliable and compressed order reduction methods out of complex physical models in different areas, such as computer graphics, control engineering, computational biology and others simulation disciplines.

In the last few years, \textit{deep autoencoder} architectures have proven successful in compressing highly complex non-linear data distributions, which arise in image and audio classification scenarios. Therefore, it is achieved a compact latent representation that encodes data variation factors in a reduced parametrization, allowing the generation of new samples that are indistinguishable from the real ones.

There is an emerging interest in applying the previous ideas to the domain of simulation, which is enhanced by the achievements made so far in recent works, \cite{kim2019deep, wiewel2019latent}. For this reason, this paper proposes the use of autoencoders as a non-linear generalization of projection-based order reduction methods, such as Proper Orthogonal Decomposition (POD).

We show, for a set of well known non-linear Partial Differential Equations (PDEs), how deep learning models
are able to find a compact basis that encode the dynamics of the original problem \eqref{general_dynamical_system}, into a small set of Ordinary Differential Equations (ODEs), in the same way that a linear wave equation can be reduced to a set of uncoupled harmonic oscillators.

\import{equations/}{general_dynamical_system}

In order to understand the capacity of neural networks to learn the global dynamics of the reduced system, we first work with phase diagrams of 2D and 3D dynamical systems. It is shown how the network is capable of learning different representative behaviours, such as dissipative dynamics, limit cycles or chaoticity. These results are then applied to make the autoencoder network learn the dynamics of reduced PDEs, proving that they can be completely solved in the hidden states of a neural network.

The great compression capacity provided by neural networks makes it possible to encode higher non-linear PDEs into 3D basis, where their dynamic behavior can be represented and their physical coherence can be easily verified.

The contribution of this work is structured in two parts. In the first part, we study classical dynamic systems representative of different phenomena. In this way, it is demonstrated that neural networks properly regularized with the loss function given in \cite{raissi2018multistep}, can learn robust approximations of a dynamic system phase space, revealing similar responses to external driving forces and identifying the bifurcations that occur due to the system parameters variation. In the second part, autoencoder neural networks are introduced as a natural non-linear extension of projection-based reduction order methods. Subsequently, it is shown how this model allows to reduce the derivative of a non-linear PDE into a compact representation, which can be integrated as a small system of ODEs. The solution can be transformed back to the original representation by using the decoded function learned by the autoencoder.

First, it is necessary to evaluate and understand the ability of a network to extrapolate and generalize the dynamics learned from a discrete set of solved trajectories; that is, to learn the time derivative function, $f(x,t)$, of the solution $x$, of the dynamical system being studied. We'll use, $\hat{x}$ to represent the coordinates in the replicated system. Next, we want to study the capacity of a latent variable neural network model to encode a high-dimensional phase space into a reduced one, in which we can calculate latent trajectories, while learning a mapping that encodes them to the original space, in a manner similar to a non-linear POD. The coordinates in the reduced replicated system are noted as $h$.

Hence,the search for a valid neural network based on a reduced order model for \eqref{general_dynamical_system}, can be carried out in two phases according to the following scheme:

\import{equations/}{problem_scheme}

In the following section, there is a collection of related work, that has served as a basis for the elaboration of this research. Following the structure indicated above, two main sections are differentiated. Each section is divided into the corresponding methods and results. The work is concluded with a brief conclusion about the results obtained.

%% file: equations/general_dynamical_system.tex
\begin{equation}\label{general_dynamical_system}
\begin{cases}
    \dfrac{\partial{u}}{\partial{t}} = F\left(u, \dfrac{\partial{u}}{\partial{x}}, \dots, \dfrac{\partial^{n}{u}}{\partial{x}^{n}}\right) \\
    \text{Boundary Conditions (BC)}\\
    \text{Initial Conditions (IC)}
\end{cases}
\end{equation}

%% file: equations/problem_scheme.tex
\begin{eqnarray}\label{problem_scheme_1}
&\text{Phase \textrm{I}:\hspace{0.2cm}}&
\begin{cases}
    \dot{x} = f(x, t) \\
    \text{IC}
\end{cases}
\qquad\Longrightarrow\qquad
\begin{cases}
    \dot{\hat{x}} = \hat{f}(\hat{x}, t) \\
    \text{IC}
\end{cases}\\
&\text{Phase \textrm{II}:\hspace{0.2cm}}&
\begin{cases}\label{problem_scheme_2}
    \partial_t{u} = f(u, \partial_x u, \dots, \partial_{x^n}u, t) \\
    \text{BC}\\
    \text{IC}
\end{cases}
\qquad\Longrightarrow\qquad
\begin{cases}
    \dot{\hat{h}} = {f}_h(h, t) \\
    \text{IC}
\end{cases}
\end{eqnarray}

%% file: sections/Related_work.tex
\section{Related work}

Regarding to the first part of the work, learning the function $f(x, t)$ leads to a certain error $\epsilon$, such that $\hat{f}(x, t) = f(x, t) + \epsilon(x, t)$. For a given domain, it can be proved \cite{funahashi1993approximation,chen1995universal} that this provides an upper bound to the error of the predicted trajectories. This, in combination with the \textit{universal approximation theorem} for neural networks \cite{cybenko1989approximation}, leads to results that establish the capability of recurrent neural networks to arbitrarily approximate any dynamical system in a bounded domain \cite{funahashi1993approximation,chen1995universal}.

About the learning of general dynamics with neural networks, \cite{tsung1995phase} shows how \textit{recurrent-feed-forward} networks can learn autonomous custom phase space portraits with different number and types of attractors. Recently, \cite{trischler2016synthesis} has extended this approach to non-autonomous systems, coupling networks that were trained separately, to accurately reproduce the dynamics of coupled systems.

The ability to learn chaotic maps is shown in \cite{navone1995learning}, which extends to the study of learning chaotic dynamical systems in \cite{bakker2000learning}. In addition, \cite{pathak2018model} uses \textit{echo-state networks} to learn higher dimensional chaotic attractors, that are present in the \textit{Kuramoto-Sivashinsky} equation.

Long term dynamic prediction is featured in \cite{pan2018long}, where a contractive loss is used to constraint the spectral radious of the jacobian matrix of the learned derivative function,thought to be linked with the unstability of the system.

A very interesting approach to get a better approximation of the system derivative was suggested in \cite{raissi2018multistep}, by building a loss function out of numerical multi-step schemes.

In the second part of the work, with the aim of synthesizing a reduced order model, as shown in the diagram \eqref{problem_scheme_2}, we face a dynamics identification problem, since the functional form of the latent dynamics is not known. In this area, \cite{brunton2016discovering} achieves great results using a dictionary sparse regression approach. It takes advantage of the fact that most dynamical systems present a sparse representation in a polynomial basis.

A POD neural network hybrid approach was introduced in \cite{wang2018model}, where Long Short Term Memory (LSTM) networks are fed with the \textit{k-component} POD solution projection, to learn the reduced dynamics. Furthermore, in \cite{vlachas2018data} high dimensional systems were solved entirely in the state-space of a LSTM network.

The most significant results achieved so far are shown in \cite{wiewel2019latent}, where convolution autoencoders coupled with LSTM in their latent space, are used to learn reduced models.

%% file: sections/PartI.tex
\section{Phase space learning}
This first part of the document corresponds to the steps described in the diagram in \ref{problem_scheme_1}.

\subsection{Methods}
Following the scheme given in \eqref{problem_scheme_1}, it is necessary to evaluate the capability of neural networks to learn phase space dynamics in bounded regions from a small set of trajectories, in most cases only one.

In the first place, it is required to define the metrics used to evaluate the quality of the replicated dynamics.

The Mean Square Error (MSE) of the differences between the ground truth and the predicted path is not a significant measure by itself, as will be explained later. We want to study the capacity of the network to preserve general qualitative features of the phase space, beyond the Least Square Errors (L2) obtained. 

We will take into account the following validation criteria:
\begin{enumerate}
    \item Robustness of the learned atractor to different initial conditions.
    \item Ability to extrapolate unseen regions of phase space.
    \item Correct response to external perturbations.
    \item Capacity to capture bifurcations in the dynamics.
\end{enumerate}

Once the evaluation criteria are defined, our goal is to obtain a machine learning algorithm to approximate the derivative, $f$, of an arbitrary dynamical system. For this it is necessary to establish the problem and select an architecture.

The machine learning problem of replicating  the system in \ref{problem_scheme_1}, is a probabilistic one. We are trying to find from the set of all  $\hat{f}(x,\theta)$ that our model can represent, the one most similar to $f$. Thus, we are not modelling a deterministic transition $X_{t+1}(X_{t})$ such as the one computed by solving an ODE, but a probabilistic one $P(X_{t+1}|X_{t})$, where the unique deterministic transitions arising from the solution of first order ODEs, are replaced by the $markovian \quad property$ of  $markov \quad   processes$, present in processes where the conditional probability distribution of future states of the process depends only upon the present state.

We are looking for a parametrized distribution $\hat{P}(f|X, \theta)$ that can be used to sample $\hat{P}(X_t\mid X_{t-1})$ with a numerical scheme. The machine learning problem is stated by requiring , $\hat{P}$, to be as close as possible to  $P(f|X)$. This means that we aim to minimize the Kullback-Leibler divergence (KL) between the two distributions, which leads to the following cost function:

\begin{equation}\label{loss_function_time_average_empiric_likelihood}
\begin{array}{c}
\displaystyle\underset{\theta}{\text{argmin }}\mathit{KL}(\theta,X)=
\underset{\theta}{\text{argmax }}\dfrac{1}{N}\sum_{k=1}^n \log(\hat{P}(\hat{f}\mid X(t_{k}),\theta))
\end{array}
\end{equation}

Since $f$ is a non observable variable, we must use it implicitly  by modelling the $P(X_{t}\mid X_{t} \dots X_{t-k},f_{t} \dots f_{t-k})$ distribution. Employing a \textit{multi-step} scheme \eqref{lmm_model}, as in \cite{raissi2018multistep}, the cost function \eqref{probability_fun_to_maximize2} is obtained.
\begin{equation}\label{lmm_model}
y_{t+s}=\sum_{i=0}^{s-1}\alpha_{i}y_{t+i}+\sum_{i=0}^{s}\beta_{i}f(y_{t+i})
\end{equation}
\begin{equation}\label{probability_fun_to_maximize2}
\underset{\theta}{\text{argmin }} \mathcal{L}(X,\theta) = \underset{\theta}{\text{argmin}}\dfrac{1}{N}\sum_{t=0}^N \log(P(X_{t+k}\mid f_{t} \dots f_{t+k},X_{t} \dots X_{t+k-1}))
\end{equation}

Finally, assuming a gaussian distribution with constant variance  $P(X_{t+k}\mid f_{t} \dots f_{t+k},X_{t} \dots X_{t+k-1})=\mathcal{N}(\mu,\sigma)$,where $\mu=\mu(f_{t} \dots f_{t+k},X_{t} \dots X_{t+k-1})$, we arrive at:
 \begin{equation}\label{final_loss_function}
 \mathcal{L}(X,\theta)=\dfrac{1}{N}\sum_{\forall t \in T}\left\|X_{t+s}-\sum_{i = 0}^{s-1}\alpha_{i}X_{t+i}-\sum_{i = 0}^{s}\beta_{i}\hat{f}(X_{t+i},\theta)\right\|^{2}_{2}
 \end{equation}

We employ a fully-connected feed-forward architecture, as in \cite{tsung1995phase,trischler2016synthesis,raissi2018multistep,pan2018long}, instead of a recurrent neural network. This is because dynamical systems can be understood as the differential deterministic equivalent of Markov processes, where $P(X_{t}\mid X_{t-1},X_{t-2}, \dots)=P(X_{t}\mid X_{t-1})$.

\begin{figure}[ht]
\centering
\includegraphics[width=0.9\linewidth]{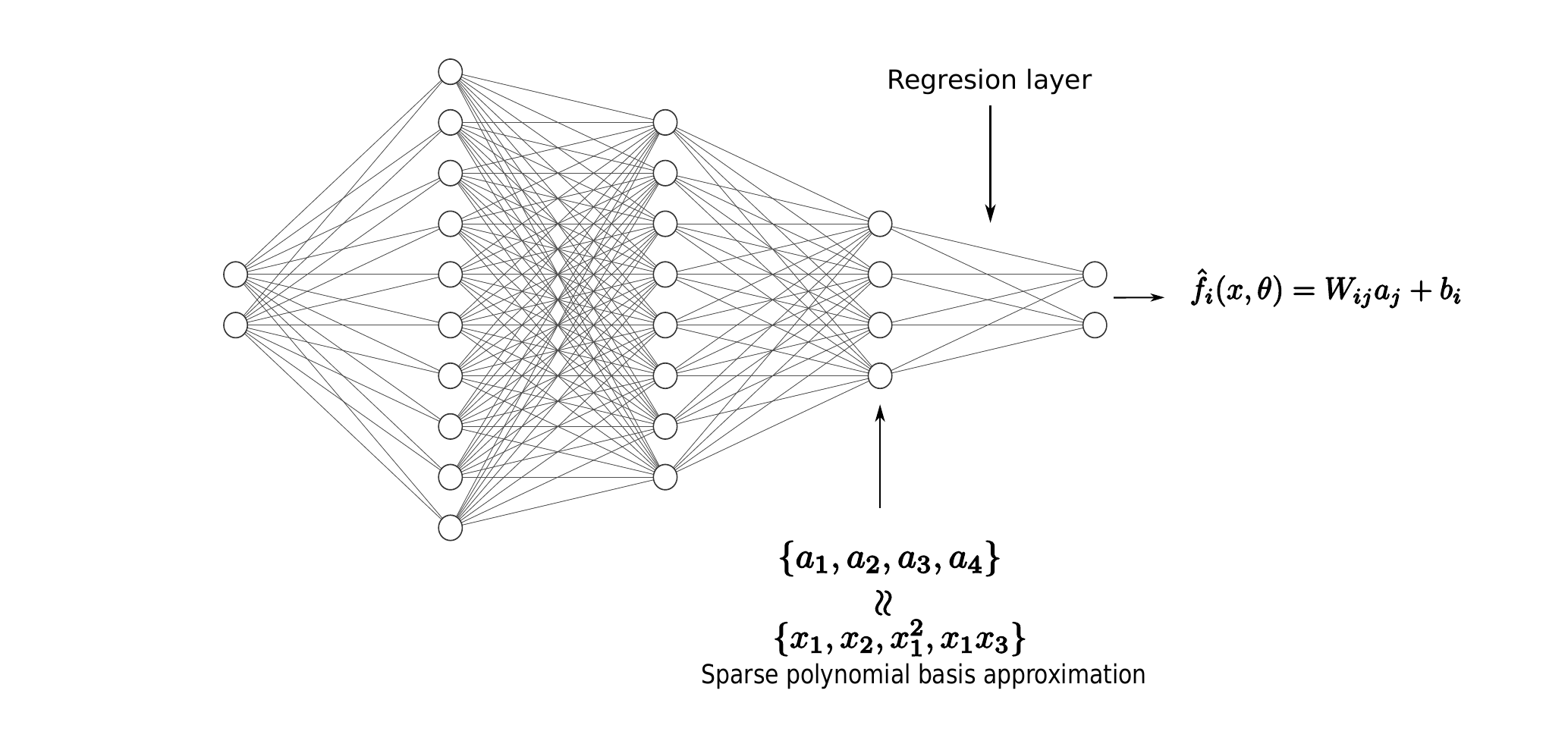}
\caption{Architecture. The last hidden layer enforces with a number of neurons equal to the number of different monomials in the system, the prior knowledge that the system is represented in a sparse polynomial basis.}
\label{psn_net}
\end{figure}

As mentioned in the previous section, all the studied systems have a derivative $f(x,t)$ with a sparse representation on a polynomial basis \cite{brunton2016discovering}. This concept is integrated into our architecture by using an encoder scheme, along with a linear regressor, which will attempt to represent the monomials present in the system, as depicted in figure \ref{psn_net}.

Conservative dynamics require a very fine tuning of learned parameters, so the error matrix, $\nabla \epsilon_{d}(x)$, should have a contractive character in order to avoid cumulative discrepances/errors. It is well known, that \textit{encoder-decoder} bottleneck architectures tend to learn contractive mappings that flow into the learnt manifold, this is, they locally behave as a linear dynamical system with $\nabla \epsilon_{d}(x)$ matrix presenting all negative eigenvalues, (or close to zero eigenvalues if we consider the associated differences equation). This character might be enhanced by certain architecture choices, such as in \textit{contractive autoencoders}~\cite{rifai2011contractive} or \textit{denoising autoencoders}~\cite{vincent2008extracting}.\\

\import{equations/}{net_error_decomposition}

\textbf{Training}\\

For all the systems studied, we compared Adam-Moulton and Backward Differentiation Formula (BDF) schemes,
with the former showing the best performance in extrapolating the dynamics of unseen regions in the phase space.

We found no evidence of the results getting worse as the order of the schemes is increased; on the contrary, the greater the order, the better the replication of the dynamics, even for higher order not $A-stable$ methods. Therefore, the BDF-6 scheme was chosen as the default scheme in the loss function.

We used Adam optimizer \cite{kingma2014adam}, with a decaying learning rate (starting in 1e-3), and a batch-size of 200. Also, Rectified Linear Unit (ReLU) \cite{nair2010rectified} activation function is utilized in all experiments.
Except for the Lorentz equation where a bigger capacity network was needed, for all the experiments in this section , a network comprised of 3 hidden layers of (40, 20, 10) neurons, plus a regression layer on top, was used.
Network weights were initialized with a He Normal Initialization \cite{he2015delving}.

\subsection{Results}
Three different simple models have been studied. The harmonic oscillator as a linear system, the Duffing equation and the Lorentz equation as examples of non-linear systems. The results obtained for each of them are described below, starting with the oscillator.\\

\import{sections/time_integration/}{linear_systems}

\import{sections/time_integration/}{Duffing}

\import{sections/time_integration/}{Lorenz_system}


%% file: equations/net_error_decomposition.tex
\textbf{Learnt system error decomposition}\\

The assessment of the learnt system error, boils down to the comparison between solutions of the two systems \eqref{net_system} with the same initial condition $x_{0}$.
\begin{equation}\label{net_system}
\begin{array}{ccc}
\begin{cases}
    \dot{\hat{x}}=\hat{f}(\hat{x}) \\
    \hat{x}_{t_{0}}=x_{0}
\end{cases}& \quad \quad &    \begin{cases}
        \dot{x}=f(x)\\
        x_{t_{0}}=x_{0}
\end{cases}
\end{array}
\end{equation}

For this purpose, a new variable is defined that quantifies the discrepancy of solutions, $z=\hat{x}-x$. Without loss of generalization, we can write $\hat{f}(x)=f(x)+\epsilon_{d}(x)$, where $\epsilon_{d}(x)$ stands for the error made at each point by the network in the calculation of the derivative.

Replacing $z=\hat{x}-x$ in \eqref{net_system} leads to the following system:
\begin{equation}
    \begin{cases}
        \dot{z}=\hat{f}(x+z)-f(x)\\
        z_{t_{0}}=0
\end{cases}
\end{equation}

At the beginning of the trajectory $z$ will be small, so we can do a power series expansion of $\hat{f}$ which will be valid for small times and will allow us to have a better understanding of the different sources or error:
\begin{equation} \label{expansion}
    \dot{z}=\hat{f}(x)+\nabla \hat{f}(x)\cdot z-f(x)=\epsilon_{d}(x)+(\nabla f(x)+\nabla \epsilon_{d}(x))\cdot z
\end{equation}

Each term in equation \eqref{expansion} has a straightforward interpretability, explicit in \eqref{zerrors}.
\begin{eqnarray} \label{zerrors}
\dot{z}_{l}&=&\epsilon_{d}(x)\nonumber\\
\dot{z}_{n}&=&\nabla f(x) \cdot z_{n}\\
\dot{z}_{s}&=&\nabla \epsilon_{d}(x) \cdot z_{s}\nonumber
\end{eqnarray}

The three equations in \eqref{zerrors} account for the different growth factors of the difference between the network prediction and the ground truth trajectory.

The local error, $z_{l}$, assess the local discrepancy that stems from the difference in the attained derivative.

The ``natural'' term, $z_{n}$, accounts for the expected divergence of close trajectories embedded in the dynamics of the system itself. In chaotic systems, the hypersensitivity to different initial conditions is going to pull trajectories apart, even if the prediction of the network is arbitrarily good. This is the main reason why we should not focus on the mean squared difference of the predicted and real trajectories to validate the quality of the prediction.

We call $z_{s}$ the stability error, because it is of the same nature as the one that appear in the stability analysis of numerical schemes, as finite differences. It states important properties about the approximation of $f(x)$ that the network should achieve. It clearly indicates that penalizing the difference $\|\hat{f}(x)-f(x)\|_{2}=\|\epsilon_{d}\|_{2}$ is not enough. Additionally, it is desirable that the eigenvalues of  $\nabla \epsilon_{d}$ are negative or small, so the network errors  $z_{s}$ are dissipative. In practice, this means the implementation of a kind of regularization penalty, which smooths the features that the network learns.\\

%% file: sections/time_integration/linear_systems.tex
\textbf{Linear systems: harmonic oscillator}\\

The general equation for the harmonic oscillator is a second order linear differential equation \eqref{linear_second_order}, where $m$, $\gamma_{m}$ and $\omega_{0m}$ are the mass, the damping coefficient and the stiffness of the oscillator, respectively.
\begin{equation}\label{linear_second_order}
m\ddot{x}+\gamma_{m}\dot{x}+\omega_{0m}x=0
\end{equation}

To attain the phase space representation of the system, we divide the last equation by $m$, incorporating it in $\gamma$, $\omega_{0}$, and then we transform it into a set of first order differential equations:
\begin{equation}\label{linear_second_order_ps}
\begin{cases}
\dot{y}=-\gamma y-\omega_{0}x\\
\dot{x}=y\\
x(t_{0})=x_{0}, y(t_{0})=y_{0}
\end{cases}
\end{equation}

The system given in \eqref{linear_second_order_ps} is particularly relevant, because it constitutes the general representation of the first order approximation to a wide range of (2D) non-linear systems in the neighbourhood of a critical point.

The best way to test the reliability of a dynamical system's reproducibility is to study its response to an external perturbation. Therefore, as indicated in \cite{trischler2016synthesis}, a sinusoidal signal of varying frequency has been added to a network trained with the homogeneous linear system \eqref{linear_second_order_ps}, being the parameters $\gamma=0.01$  and $\omega_{0}=1$.

Subsequently, it has been compared with the integrated numerical solution of the perturbed system. The resulting system corresponds to a forced damped harmonic oscillator:
\begin{equation}\label{linear_second_order_ps_forced}
\begin{cases}
\dot{y}=-\gamma y-\omega_{0}x+g(t)\\
\dot{x}=y\\
x(t_{0})=x_{0},y(t_{0})=y_{0}
\end{cases}
\end{equation}

The solution of an harmonically driven damped oscillator is an harmonic oscillation with an amplitude dependent on the frequency of the driving signal, showing a resonance peak near the natural frequency $\omega_{0}$.

\begin{figure}[H]
\centering

\begin{subfigure}[h]{0.18\linewidth}
\includegraphics[width=\linewidth]{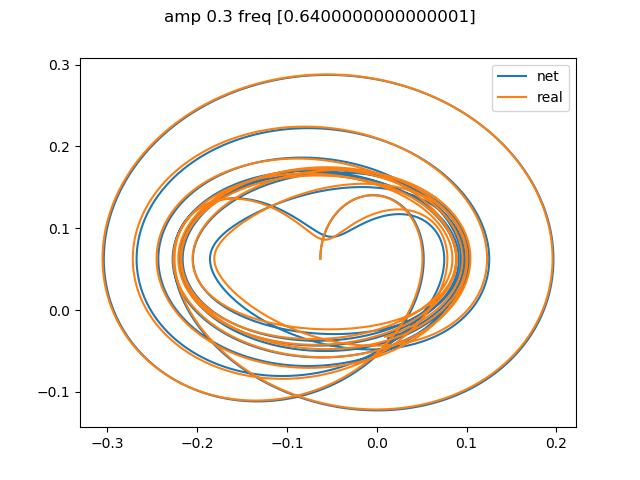}
\end{subfigure}
\begin{subfigure}[h]{0.18\linewidth}
\includegraphics[width=\linewidth]{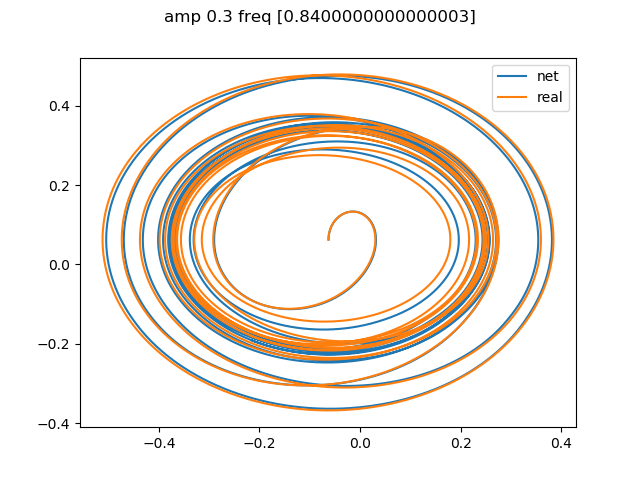}
\end{subfigure}
\begin{subfigure}[h]{0.18\linewidth}
\includegraphics[width=\linewidth]{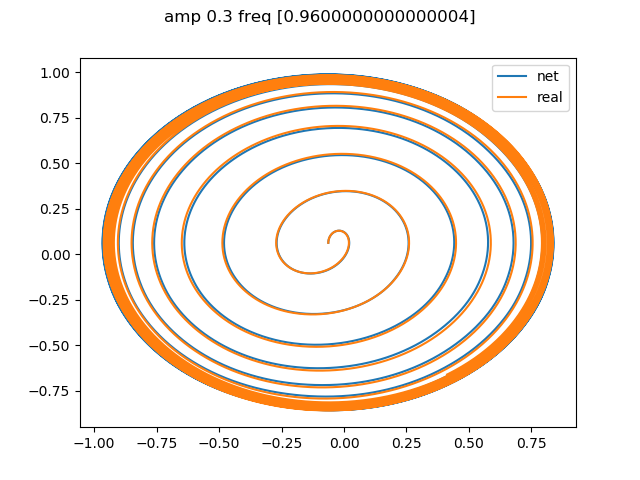}
\end{subfigure}
\begin{subfigure}[h]{0.18\linewidth}
\includegraphics[width=\linewidth]{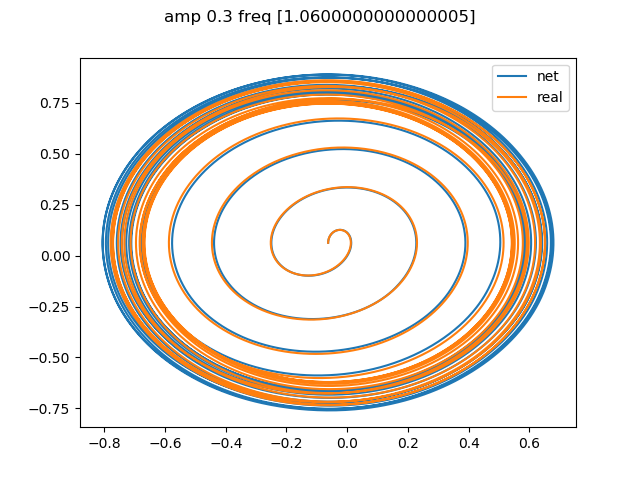}
\end{subfigure}
\begin{subfigure}[h]{0.18\linewidth}
\includegraphics[width=\linewidth]{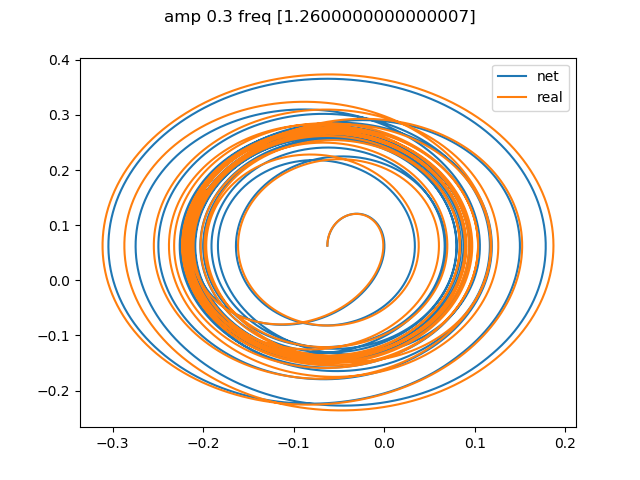}
\end{subfigure}

\begin{subfigure}[h]{0.18\linewidth}
\includegraphics[width=\linewidth]{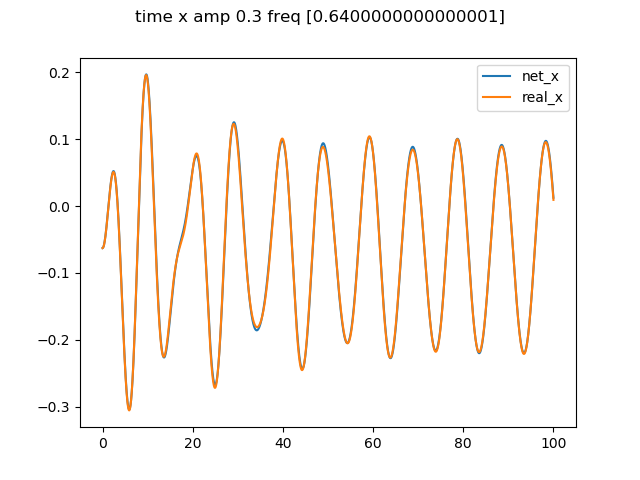}
\end{subfigure}
\begin{subfigure}[h]{0.18\linewidth}
\includegraphics[width=\linewidth]{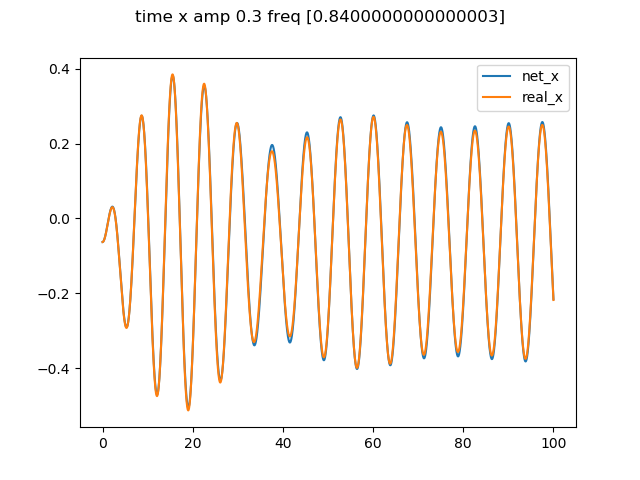}
\end{subfigure}
\begin{subfigure}[h]{0.18\linewidth}
\includegraphics[width=\linewidth]{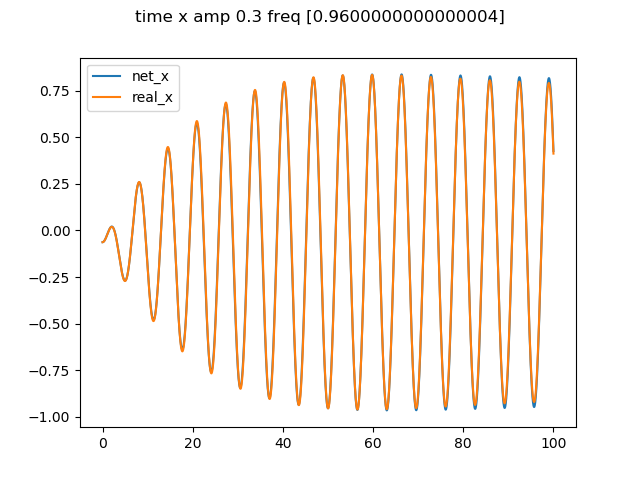}
\end{subfigure}
\begin{subfigure}[h]{0.18\linewidth}
\includegraphics[width=\linewidth]{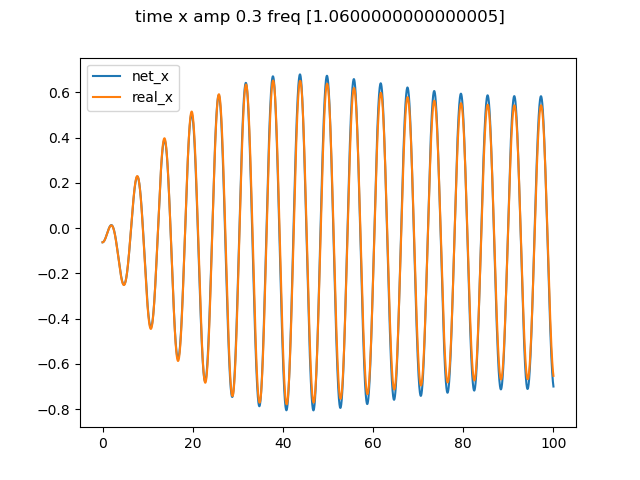}
\end{subfigure}
\begin{subfigure}[h]{0.18\linewidth}
\includegraphics[width=\linewidth]{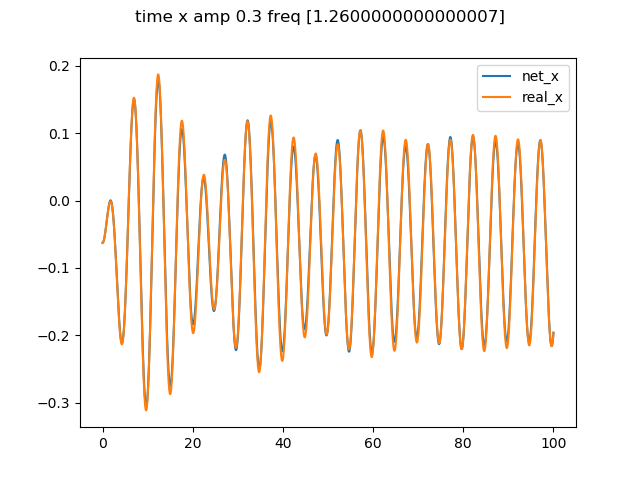}
\end{subfigure}

\caption{Solution of an harmonically driven damped oscillator. Neural network performance in orange, numerical solution in blue.}
\label{forced_results}
\end{figure}

In figure \ref{forced_results}, it is shown how similar the response of both systems is, presenting a resonance peak, ( driving force frequency that causes the maximum amplitude), virtually at the same driving frequency.\\

%% file: sections/time_integration/Duffing.tex
\textbf{Duffing equation}

Duffing equation is a non-linear second-order differential equation, given by:
\begin{equation}\label{duffing}
\ddot{x}+ \delta \dot{x}+ \alpha x+  \beta x^{3} =\gamma \cos{\omega t}
\end{equation}

Equation \eqref{duffing}, is suited for modelling oscillators with non-linear, odd restoring forces. \eqref{duffing_system}.
\begin{equation}\label{duffing_system}
    \begin{cases}
    \dot{x}=y\\
    \dot{y}=-\delta{y}-\alpha{x}-\beta x^{3}+\gamma \cos{\omega t}\\
    x(t_{0})=x_{0}, y(t_{0})=y_{0}
    \end{cases}
\end{equation}

\begin{figure}[H]
\centering
\begin{subfigure}[h]{0.23\linewidth}
\includegraphics[width=\linewidth]{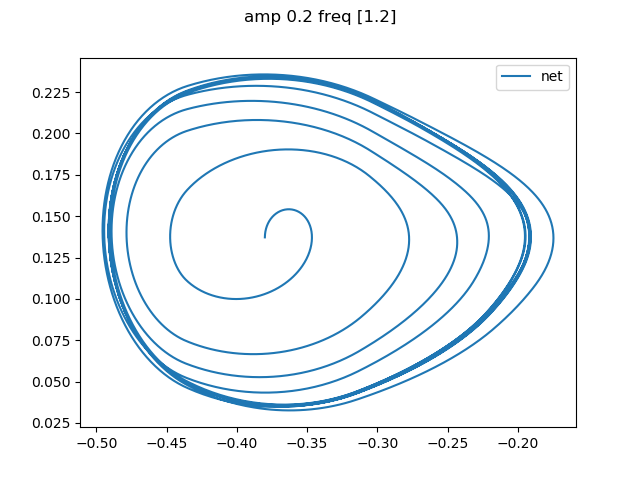}
\end{subfigure}
\begin{subfigure}[h]{0.23\linewidth}
\includegraphics[width=\linewidth]{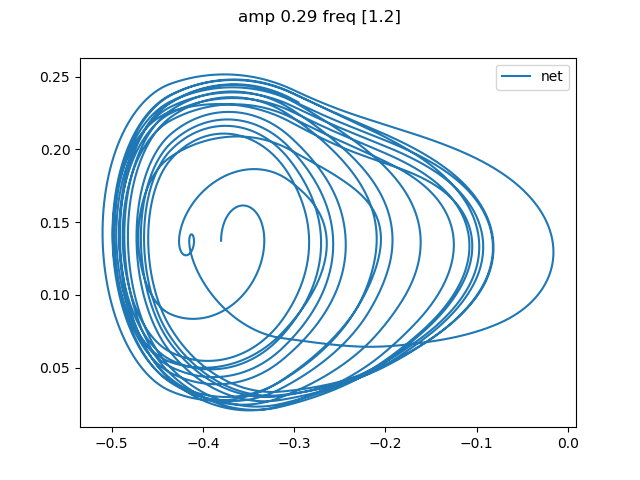}
\end{subfigure}
\begin{subfigure}[h]{0.23\linewidth}
\includegraphics[width=\linewidth]{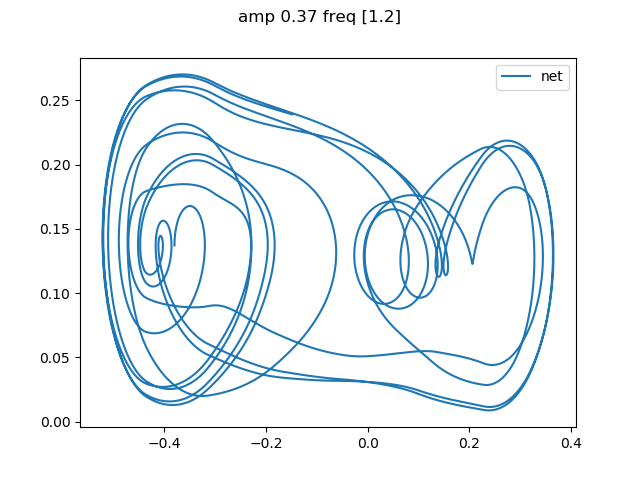}
\end{subfigure}
\begin{subfigure}[h]{0.23\linewidth}
\includegraphics[width=\linewidth]{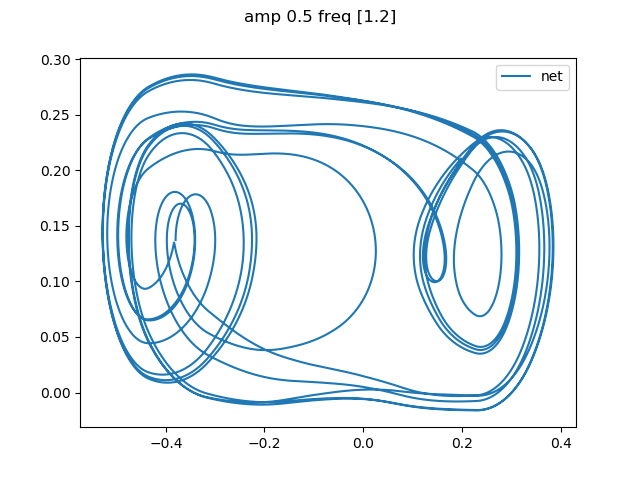}
\end{subfigure}

\begin{subfigure}[h]{0.23\linewidth}
\includegraphics[width=\linewidth]{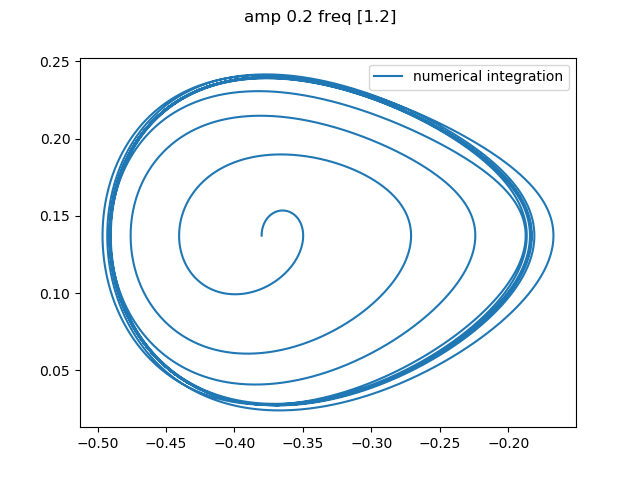}
\end{subfigure}
\begin{subfigure}[h]{0.23\linewidth}
\includegraphics[width=\linewidth]{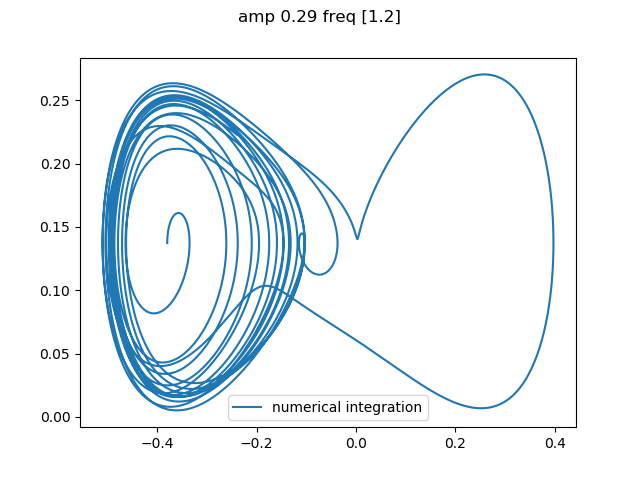}
\end{subfigure}
\begin{subfigure}[h]{0.23\linewidth}
\includegraphics[width=\linewidth]{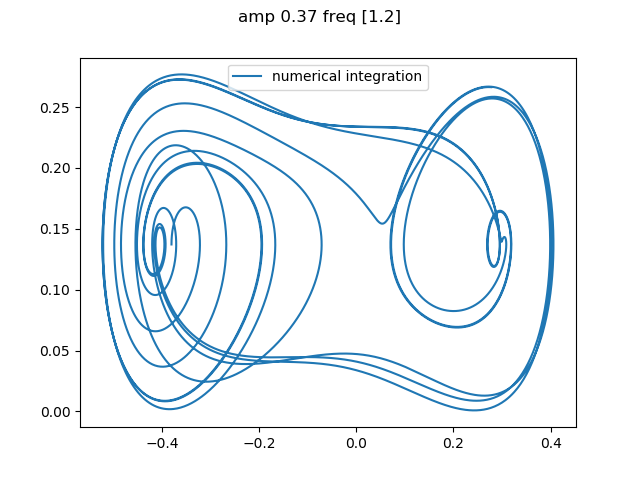}
\end{subfigure}
\begin{subfigure}[h]{0.23\linewidth}
\includegraphics[width=\linewidth]{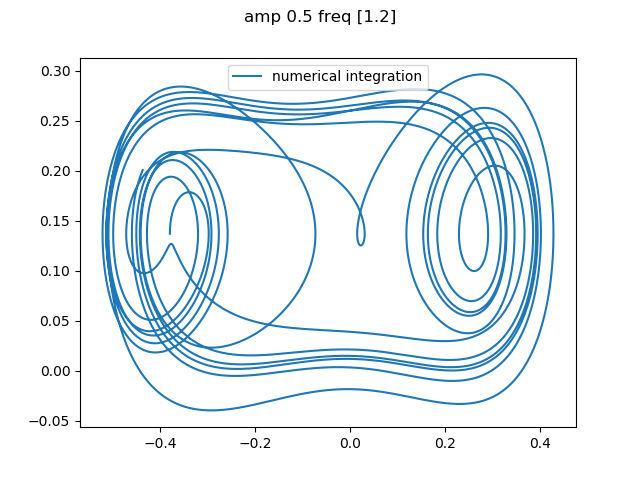}
\end{subfigure}
\caption{Solution of an oscillator with non-linear odd restoring forces. The first row corresponds to a forced network trained on the homogeneous Duffing equation. The second one shows the ground truth forced system.}
\label{duffing_fig}
\end{figure}

By using the parameters $\gamma=0$, $\alpha=-1$, $\beta=1$, $\delta=0.3$ and $\omega=1.2$, the potential $V=\frac{1}{2}\alpha x^{2}-\frac{1}{4}\beta x^{4}$, that generate the restoration force $-\nabla V=-\alpha x -\beta x^{3}$, is a two well potential. Without any driving force, the system has two fixed point atractors, that correspond to a particle being trapped in one of the two wells, as shown in figure~\ref{duffing_fig}.

As the amplitude of the driving force, $\gamma$,  increases, the particle is more capable of jumping from one well to another. Once a certain amplitude is exceeded, a series of \textit{period-doubling} bifurcations take place, so the system's behavior becomes chaotic. 

Proceeding in an analogous manner to what was done with the damped driven oscillator, we first train the network with a trajectory calculated solving the \eqref{duffing_system} system without the $\gamma cos(\omega t)$ term. Once the network is trained we add the driving term to the model output and we sweep across a range of  $\gamma$ values from  $0$ to $0.6$, knowing that the bifurcation takes place near $\gamma=0.3$.

As it can be seen in the figure \ref{duffing_fig}, both the neural network system and the original one, suffer a transition towards chaos at approximately the same value of $\gamma$. The trajectories of both systems differ considerably after some time. This is not an indicator of low accuracy, but rather the expected behavior of a chaotic system, whose arbitrarily close starting points are carried away by positive eigenvalues of $\nabla f$, as indicated in the error description part.

For $\gamma$ values well above the bifurcation, the chaotic behavior starts to decrease and the decorrelation times of both the network and the original systems increase. This is due to the fact that the \textit{Maximum Characteristic Lyapunov exponent} of the system decreases to a point where the dynamics are no longer chaotic.\\

%% file: sections/time_integration/Lorenz_system.tex
\textbf{Lorenz equation}\\

The Lorentz system \eqref{Lorenz_system} is comprised of a set of three non-linear ODEs, and is frequently used as the paradigmatic system for the exemplification of chaotic solutions. It was developed  as a simplified mathematical model for atmospheric convection, by truncating a series expansion of the \textit{Navier-stokes} equations, describing a two-dimensional fluid layer uniformly warmed from below and cooled from above.

\begin{equation}\label{Lorenz_system}
    \begin{cases}
   \dot{x}=\sigma(y-x)\\
   \dot{y}=x(\rho-z)-y\\
   \dot{z}=xy-\beta z
    \end{cases}
\end{equation}

\begin{figure}[H]
\centering
\includegraphics[width=0.4\linewidth]{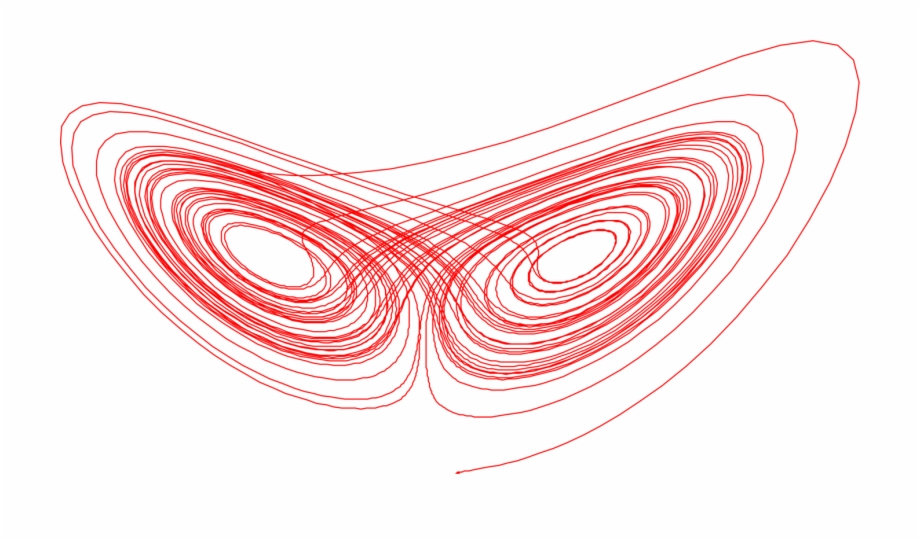}
\caption{Lorentz's atractor.}
\label{lorenzbut}
\end{figure}

 Using the values $\sigma=10$, $\beta=8/3$ and $\rho=28$, the systems presents chaotic solutions, where trajectories are pulled toward a fractal set that constitutes Lorentz's strange atractor, figure~\ref{lorenzbut}. This attractor presents global stability, in the sense that almost all points in the phase space are attracted to it. Once in the attractor, any point will spiral around one of the fixed points before shooting at the other symmetric one. The time it will spend on each of the wings, or which one the point will go to first, depends mostly on the initial conditions, which, although deterministic, are computationally unpredictable. This arises from the deep entanglement of the stable and unstable manifolds of the different fixed points \cite{osinga2002visualizing}. Preserving this fine structure of phase space, needed to replicate chaotic behaviours, means that the network reconstruction of $f$ should be much more accurate, hence, more capacity will be required.

\begin{figure}[h]
\centering
\begin{subfigure}[h]{0.31\linewidth}
\includegraphics[width=\linewidth]{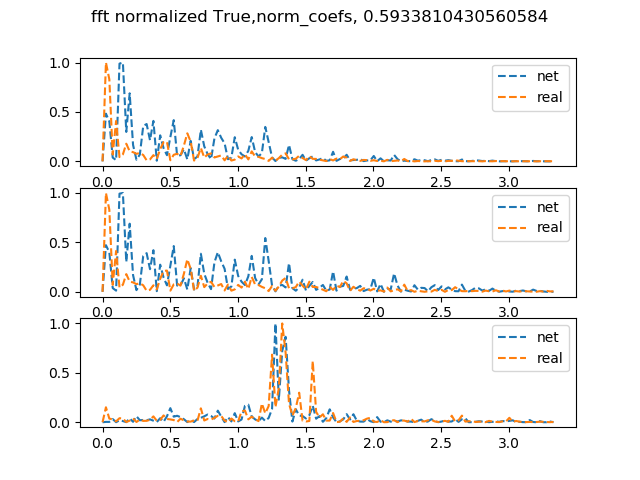}
\end{subfigure}
\begin{subfigure}[h]{0.31\linewidth}
\includegraphics[width=\linewidth]{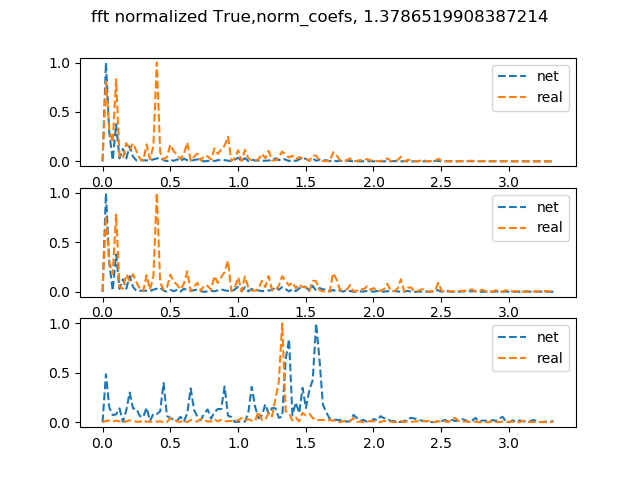}
\end{subfigure}
\begin{subfigure}[h]{0.31\linewidth}
\includegraphics[width=\linewidth]{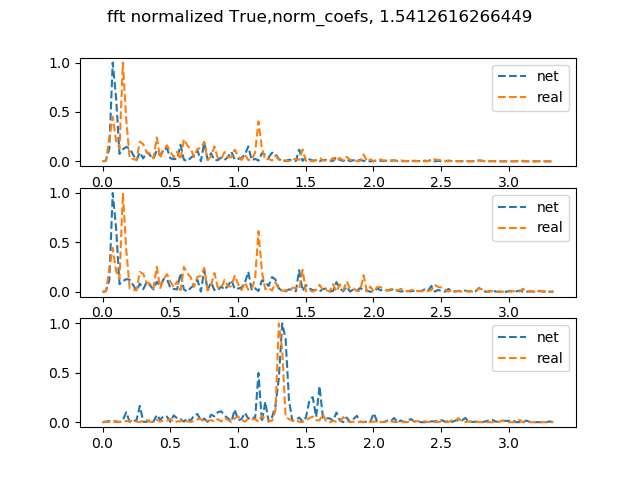}
\end{subfigure}

\begin{subfigure}[h]{0.31\linewidth}
\includegraphics[width=\linewidth]{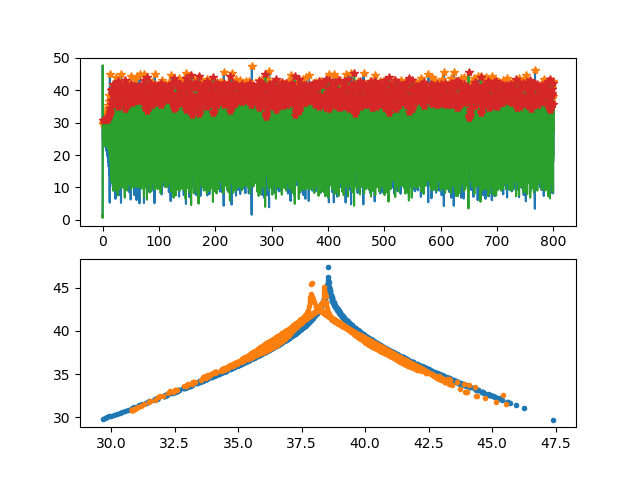}
\end{subfigure}
\begin{subfigure}[h]{0.31\linewidth}
\includegraphics[width=\linewidth]{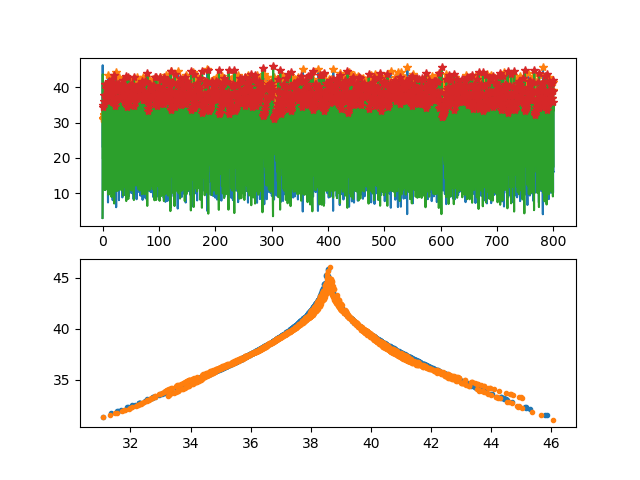}
\end{subfigure}
\begin{subfigure}[h]{0.31\linewidth}
\includegraphics[width=\linewidth]{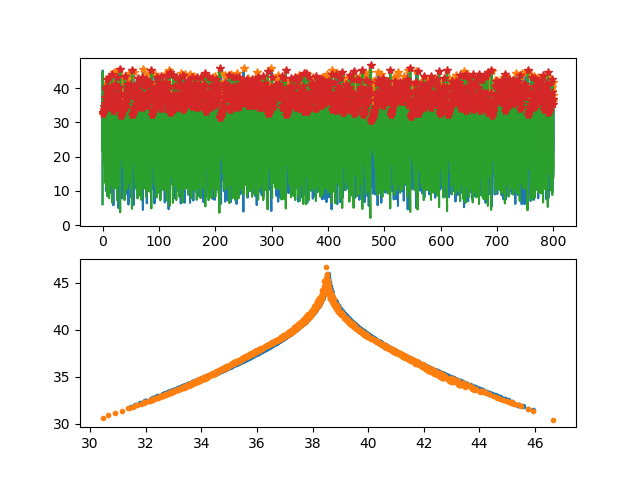}
\end{subfigure}

\begin{subfigure}[h]{0.31\linewidth}
\includegraphics[width=\linewidth]{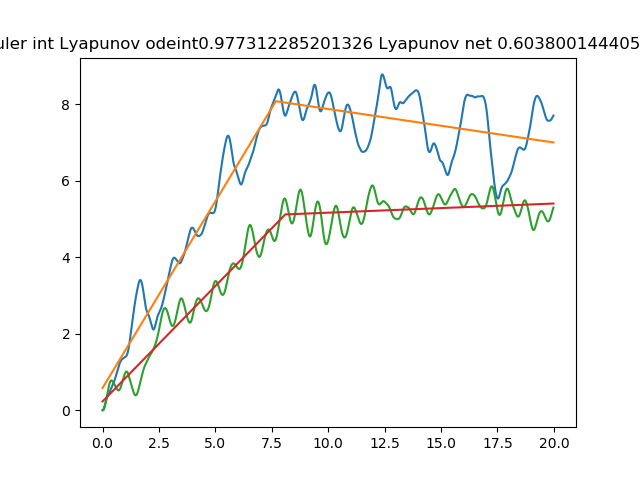}
\end{subfigure}
\begin{subfigure}[h]{0.31\linewidth}
\includegraphics[width=\linewidth]{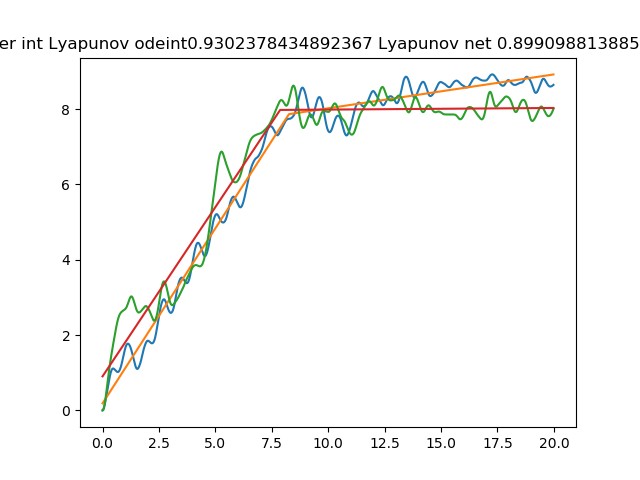}
\end{subfigure}
\begin{subfigure}[h]{0.31\linewidth}
\includegraphics[width=\linewidth]{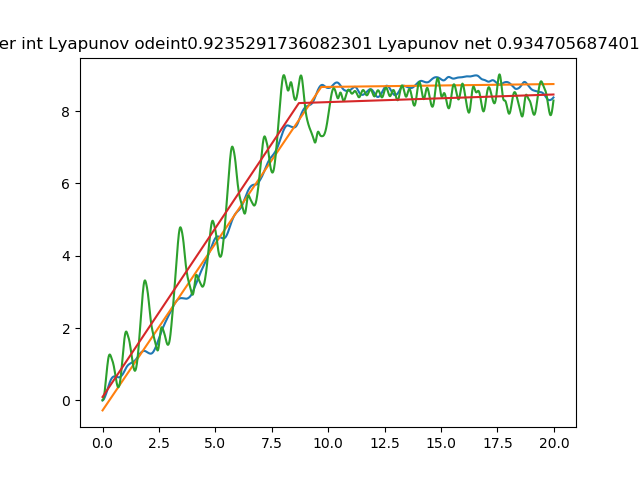}
\end{subfigure}

\caption{Fourier transform, Lorenz maps and $MCLE$ estimations   of a neural network with 3 hidden layers and 16 neurons (left column), 3 hidden layers and 30 neurons (middle column) and 2 hidden layers and 60 neurons (right column).Solution of the ODE in blue, approximated solution in orange. We can see how the spectral power composition of the solutions is better matched in the case of the higher capacity networks, while the small one fails at capturing higher frequencies.$MCLE$ is approximated by the slope of the average $log(distances)$ of initially close trajectories ( the flattening part of the curve takes place after going through the Lorenz butterfly center, that acts as a bottleneck that $grinds$ initial correlations).We can see the close match of slopes between the curves of the approximated model and the original one.}
\label{Lorentz_fig}
\end{figure}

To obtain a reliable chaotic replication of the attractor, 100 neurons have been used in the first hidden layer. In this case, the Lorentz map, the Maximum Characteristic Lyapunov Exponent $(MCLE)$ estimation and the Fourier transform of the trajectories agree on the chaotic nature of the learned attractor, as can be seen in the figure \ref{Lorentz_fig}.

By increasing the depth (number of hidden layers) of the network while maintaining the width (number of neurons), chaotic dynamics are achieved with a reduced number of parameters. This is because the number of regions that can be separated by a neural network grows exponentially with depth \cite{montufar2014number}. In addition, because the two fixed points are symmetrical, there are features that can be shared and do not need a separate parameterization, as it happens in a shallow architecture.

\begin{figure}[h]
\centering
\begin{subfigure}[h]{0.48\linewidth}
\includegraphics[width=\linewidth]{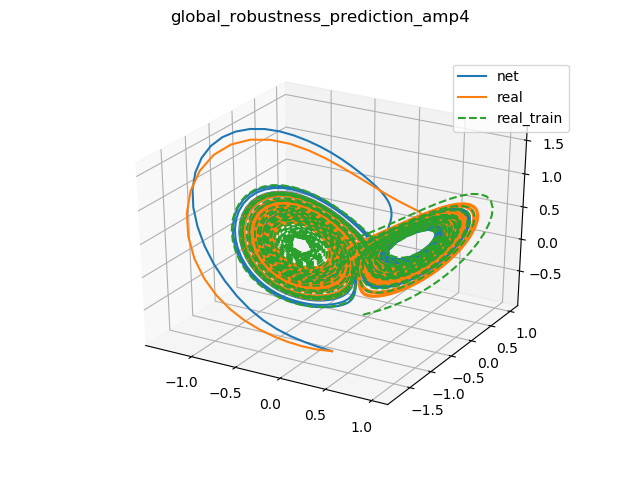}
\end{subfigure}
\begin{subfigure}[h]{0.48\linewidth}
\includegraphics[width=\linewidth]{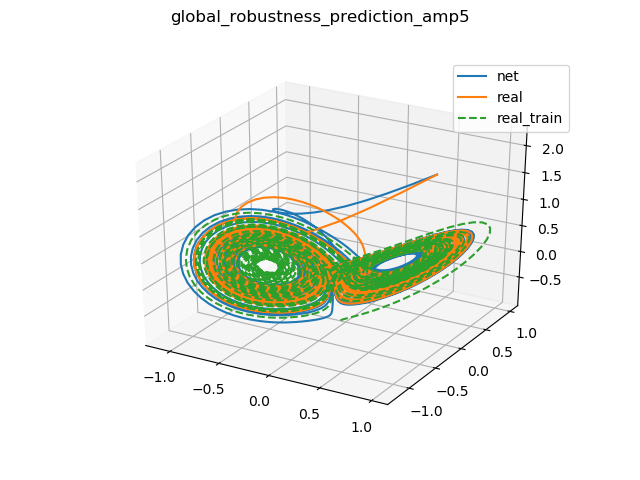}
\end{subfigure}
\begin{subfigure}[h]{0.48\linewidth}
\includegraphics[width=\linewidth]{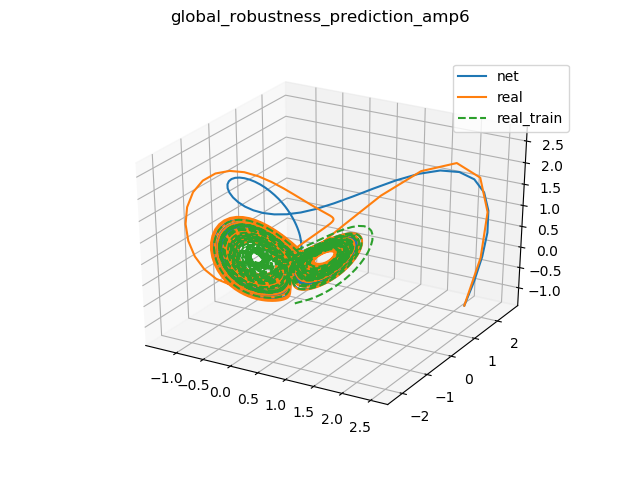}
\end{subfigure}
\begin{subfigure}[h]{0.48\linewidth}
\includegraphics[width=\linewidth]{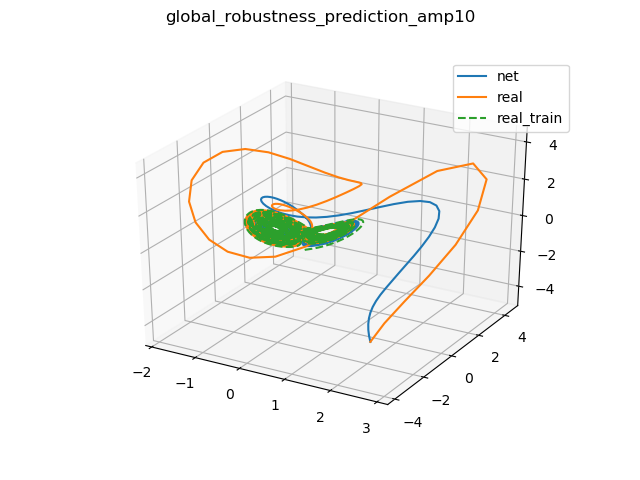}
\end{subfigure}
\caption{Different trajectories for a neural network with 60 neurons and 2 hidden layers. Distant points are pulled back to the atractor. The training trajectory in green, the neural network results in blue and test trajectory in orange.}
\label{60_neurons_net}
\end{figure}

In figure \ref{60_neurons_net} we use a neural network with 60 neurons in the first layer, 40 in the second, 4 in the third and 3 in the linear regression layer. It is shown how the global stability of the learnt atractor is preserved far from the training trajectory, promoting even further the perspective of deep neural networks as dynamical systems.

Another advantage of deep learning is the ability to learn bifurcations. By maintaining the value of $\sigma$ and $\beta$ parameters, but changing $\rho$, huge variations in system dynamics are observed. This structural modification is caused by changes in the stability of the fixed points and the appearance of new ones. The study of this phenomenon is included in the theory of bifurcation. Here is the behaviour for some significant values:
\begin{enumerate}
    \item If $0<\rho<1$, the origin is the only equilibrium point, towards which all the orbits converge.
    \item If $\rho=1$, a \textit{supercritical pitchfork} bifucation occurs, the origin become unstable and a pair of stable symmetric fixed points appear.
    \item If $\rho=\frac{\sigma(\sigma+\beta+3)}{\sigma-\beta-1}\approx 24.73$, the eigenvalues of both symmetric fixed points cross the complex plane and lose their stability through a subcritical \textit{Hopf bifurcation}~\cite{hirsch2012differential}.
\end{enumerate}

\par As done in \cite{raissi2018multistep}, learning the bifurcations can be re-stated as learning the phase space of the system \eqref{Lorenz_system_parametric}. Where, $\rho$ is treated as an additional variable, whose fixed character is described by $\dot{\rho}=0$.

\import{equations/}{lorenz_parametric.tex}

\begin{figure}[h]
\centering
\begin{subfigure}[h]{0.31\linewidth}
\includegraphics[width=\linewidth]{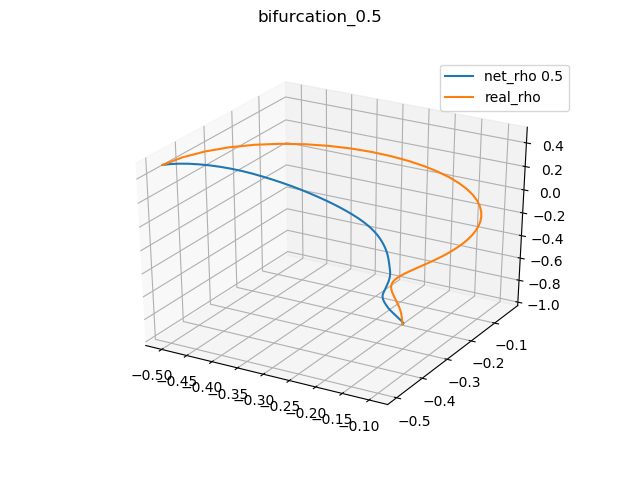}
\end{subfigure}
\begin{subfigure}[h]{0.31\linewidth}
\includegraphics[width=\linewidth]{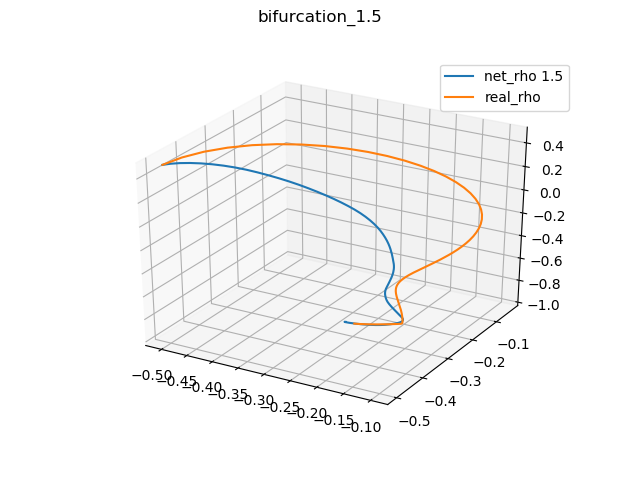}
\end{subfigure}
\begin{subfigure}[h]{0.31\linewidth}
\includegraphics[width=\linewidth]{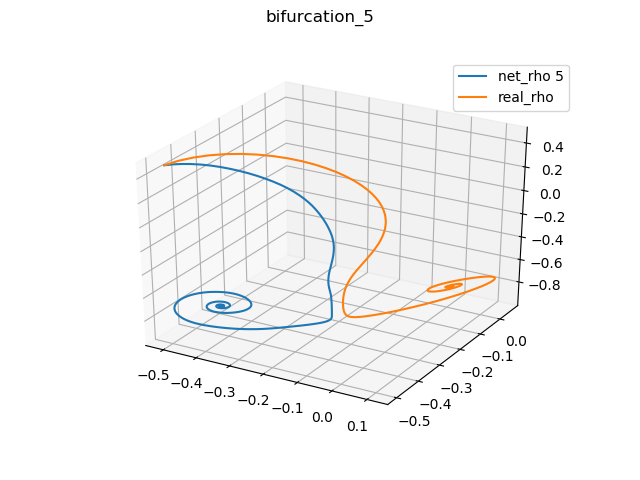}
\end{subfigure}
\begin{subfigure}[h]{0.31\linewidth}
\includegraphics[width=\linewidth]{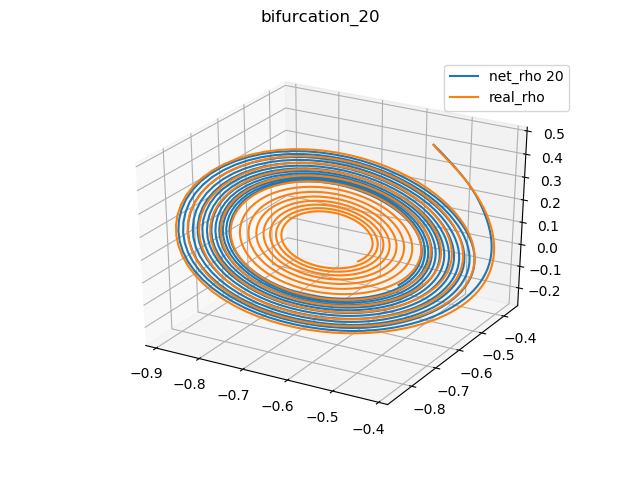}
\end{subfigure}
\begin{subfigure}[h]{0.31\linewidth}
\includegraphics[width=\linewidth]{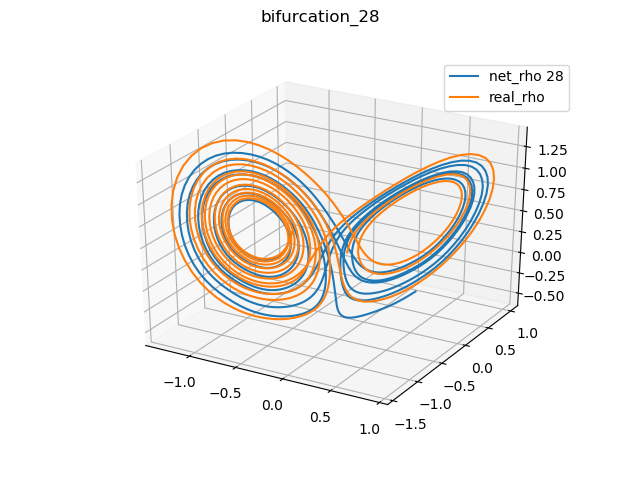}
\end{subfigure}
\begin{subfigure}[h]{0.31\linewidth}
\includegraphics[width=\linewidth]{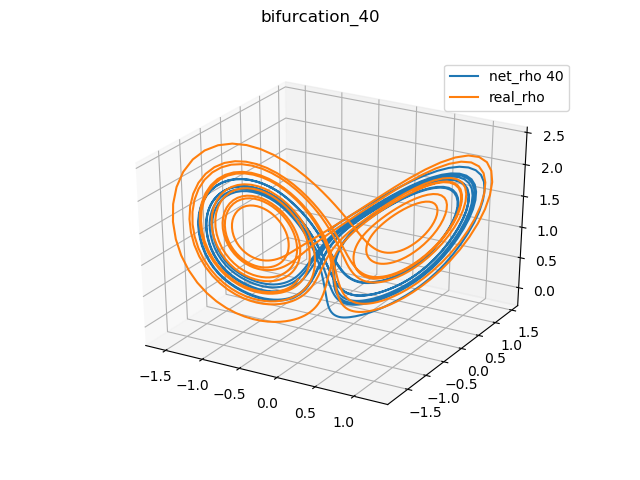}
\end{subfigure}
\caption{Network predictions and numerical integration results for different values of $\rho$.}
\label{lorentz_bifurcations}
\end{figure}

A network with $80$, $60$, $10$ and $4$ neurons in the hidden layers has been trained with $\rho\in(1\dots20)$, so that the network does not see data from orbits in regimes that correspond to $\rho<1$ or $\rho>24.7$. For the evaluation, the network is fed with the desired value of $\rho$ and the conditioned trajectory is integrated as in the previous experiments. As shown in figure \ref{lorentz_bifurcations}, the network is able to capture both non seen bifurcations at $\rho=1$ and $\rho=24.7$.

%% file: equations/lorenz_parametric.tex
\begin{equation}\label{Lorenz_system_parametric}
    \begin{cases}
                   \dot{x}=\sigma(y-x)\\
               \dot{y}=x(\rho-z)-y\\
               \dot{z}=xy-\beta z\\
               \dot{\rho}=0\\
               \vec{x}(t_{0})=\vec{x_{0}},\rho(t_{0})=\rho_{0}

    \end{cases}
\end{equation}

%% file: sections/PartII.tex
\section{Latent phase space learning}
\begin{figure}[H]
\centering
\includegraphics[width=1\linewidth]{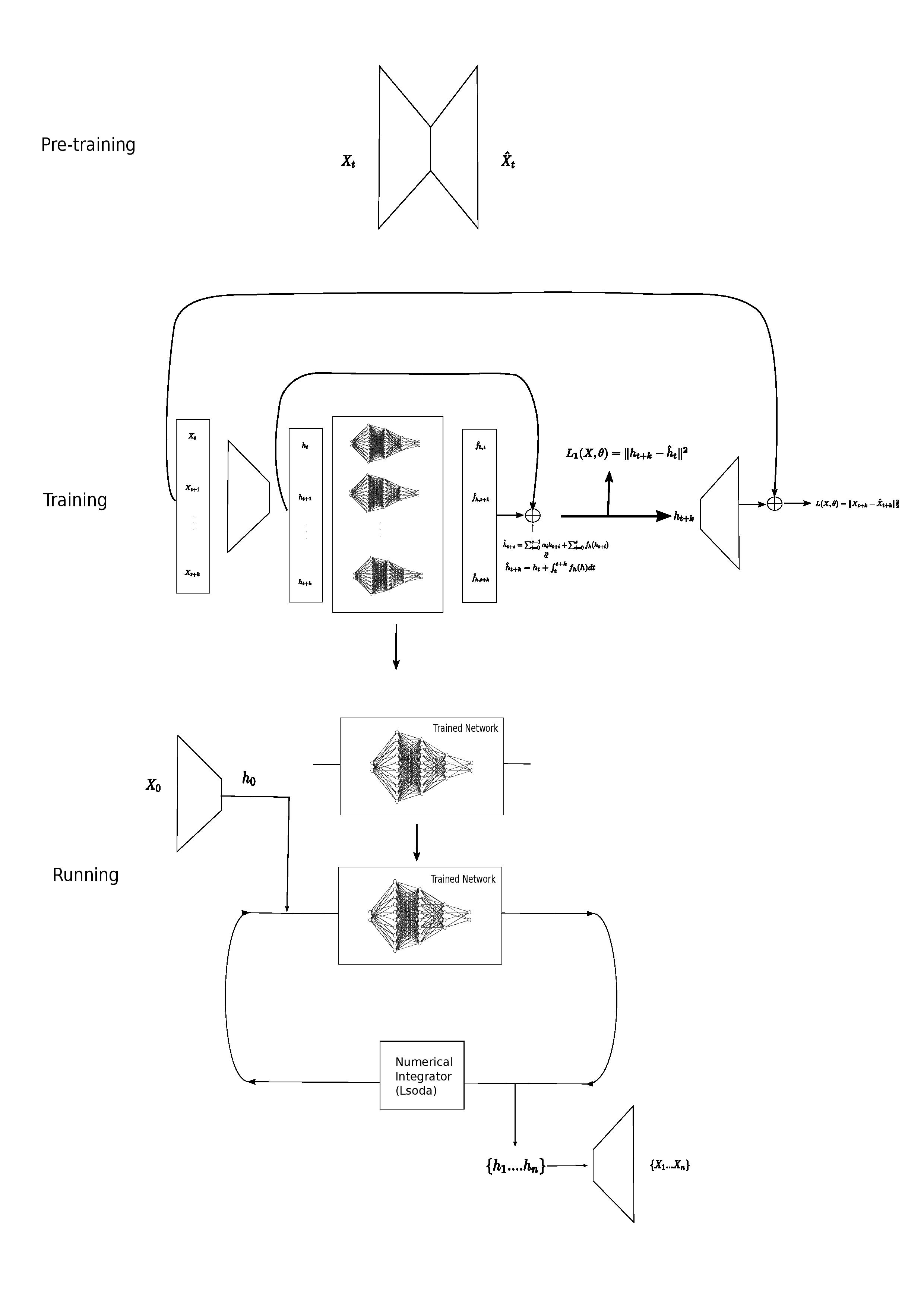}
\caption{Autoencoder and latent integration, training and running scheme}
\label{training scheme}
\end{figure}

This second part of the document corresponds to the steps described in the diagram in \ref{problem_scheme_2}.
The methods and the training process described below, are schematically displayed in figure \ref{training scheme}.

\subsection{Methods}

Once we have studied simpler models, in this second part of the work, we will move forward to deal with dynamical systems described by PDEs \eqref{general_dynamical_system}, with a large number of degrees of freedom. These can be reduced by searching for a suitable coordinate change, which better represents the structure of the latent phase space. 
The first training stage will consists of an unsupervised learning stage, where we aim to capture the manifold where the trajectory exists,this is finding an initial basis to integrate the latent dynamical system, this step involves the minimization of the reconstruction loss term \ref{latent_probability_fun_to_maximize_pre}.

The probability distribution that we are modelling now is $P(X_{t+k}|X_{t},\theta)$, and we want to factorize it into a projection to a reduced coordinate system $P(h_{t}|X_{t})$ and  a latent integration part $P(h_{t+k}|h_{t})$, this factor would play the role of $P(x_{t+k}|x_{t})$ in last section. Under the assumption of normal errors of constant variance, this factorization leads us to  $P(X_{t+k}|X_{t},\theta)=P(X_{t+k}|h_{t+k})P(h_{t+k}|h_{t \dots t+k-1},f_{t},X_{t},\theta)=\mathcal{N}(\mu_{1}=\hat{X}_{t+k}(h_{t+k},\theta_{encoder}),\sigma)\mathcal{N}(\mu_{2}=\hat{h}_{t+k}(\theta_{integration},h_{t}(X_{t},\theta_{decoder})),\sigma)$. With this probability distribution we can lay out the optimization problem in the same way as in part I, as it is shown in \eqref{latent_probability_fun_to_maximize3}. 

\begin{eqnarray}\label{latent_probability_fun_to_maximize_pre}
\displaystyle\underset{\theta}{\text{argmax }}\mathcal{L}(X,\theta) &=&\underset{\theta}{\text{argmax}}\dfrac{1}{N}\sum \log(P(X_{t}| \hat{X_{t}})\nonumber\\ &=& \underset{\theta}{\text{argmin }}\dfrac{1}{N}\sum \left\|X_{t}-\hat{X}_{t}(h_{t}(\theta,X_{t})\right\|^{2}
\end{eqnarray}
\begin{eqnarray}\label{latent_probability_fun_to_maximize3}
\underset{\theta}{\text{argmax }}\mathcal{L}(X,\theta)=
\displaystyle\underset{\theta}{\text{argmax } }\dfrac{1}{N}\sum_{k=1}^{n}log(P(X_{t+k}|h_{t+k})&\nonumber\\P(h_{t+k}| h_{t+1}\dots h_{t+k-1},f_{ht+1}\dots f_{ht+k},X_{t+1}\dots X_{t+k-1}))&
\end{eqnarray}

Inserting the multi-step approximation of the latent derivative in the latent space of the autoencoder to model $P(h_t+k|h_{t})$, we get the cost function \eqref{latent_probability_fun_to_maximize4}. The autoencoder and the derivative parameters are denoted $\theta_{\textit{aut}}$ and $\theta_{\textit{der}}$, respectively.
\begin{eqnarray}\label{latent_probability_fun_to_maximize4}
\mathcal{L}(X,\theta) &=& \sum_{k=1}^{n}\left\|X_{t+k}-\hat{X}_{t+k}(h_{t+k})\right\|^{2}\nonumber\\ &+& \left\|h_{t+k}-\sum_{i=0}^{k-1}\alpha_{i}h_{t+i}(X,\theta_{\mathit{aut}})-\sum_{i=0}^{k}f_{ht+i}(h,\theta _{\mathit{der}})\right\|^{2}
\end{eqnarray}

\textbf{Training}\\

Training and evaluation is carried out by following these steps:
\begin{enumerate}
    \item In a pre-training stage, $X_{t}$ is reconstructed.
    \item In the training stage, we insert the derivative network studied in the first part of this work, that models $P(f_h|h)$.
    \item In the evaluation stage, we unplug the trained derivative network, encode an initial condition $X_{0}\rightarrow h_{0}$ and integrate the derivative network in the latent space. We later decode this integrated solution and recover the $X$ representation of the trajectory.
\end{enumerate}

For the autoencoder, it was used a 100 neuron and 9 hidden layers network with ReLU activations, except for the latent and regressor layers. It was used a width factor between layers of $3/4$, and a $60$ ReLU, $40$ ReLU, $20$ linear network to model the latent derivative.

A denoising autoencoder \cite{vincent2008extracting} architecture was fundamental to learn useful robust features. A corrupted noise of amplitude $10^{-3}$ was injected in the normalized input data.

Adam optimization algorithm \cite{kingma2014adam} was used in the \textit{backpropagation} optimization to train the network, along with a $0.1$ factor decaying learning rate every $200$ epochs and a batch-size of $200$ samples. Smaller networks would be enough to successfully learn most of the problems, but for consistency sake, the architecture was kept fixed.

\section{Results}
The PDE training data was acquired by solving different equations with the MATLAB spectral solver package \texttt{Chebfun}. We used $512$ nodes and periodic boundary conditions in all cases, down-sampled to $64$ to train the network.

When solving equations using a spectral method with this resolution, some numerical artifacts may occur. This is the case of the formation of shock waves in the \textit{Burgers equations}. High frequency modes, which cannot be represented in discretization, give rise to phenomena like the Gibbs one. As will be seen below, this is not a problem for a neural network, since it achieves a more compact and richer representation than the spectral one.

The integration time for each experiment was set in order to clearly manifest the characteristic dynamics given by each equation with every coefficient set to 1. Burger's equation integration time should capture the formation and damping of the shock wave. In the Korteweg-de-Vries equation, the time used captured 3-4 cycles of soliton (semi)periodic movement. In the Kuramoto Sivashinsky equation, the time was set to ensure fully developed turbulent behaviour, where the intermittency characteristics of the turbulent patterns where shown.

Spatial domain bounds in all plots were scaled to  $[-1,1]$. The spatial domain for the Burger's equation was $[-8,8]$,with an initial condition of  $e^{-(x+2)^{2}}$ and integration time of $70$. For the Korteweg-de-Vries equation, in the single soliton case, the spatial domain is $[-10,10]$, with an initial condition $e^{-(x-7)^{2}/10}$ and an integration time of $12$. For the multiple solition case we used a domain of $[-20,20]$ and an initial condition $sin(\frac{\pi x}{20})$. With Kuramoto Sivashinsky equation, the spatial domain was $[50,0]$ and the integration time was $250$, with the initial condition $cos(\frac{x}{8})(1+sin(\frac{x}{8}))$.

The  temporal discretization used is problem dependent. We  observed that, in general, the fitted networks are able to preserve numerical stability with temporal deltas of greater magnitude than in the case of the spectral method integration, so, in every experiment, with the integration time fixed, we sought for the biggest $dt$ that made the spectral-method integration convergent, and then downsampled it by a factor of 2 to feed the network, we saved $25\%$ of the last time samples for testing and used the rest for training.\\

\textbf{Viscid Burgers}\\

Burger's equation \eqref{burger_eq} is a simplification of  the incompressible Navier-Stokes equation. It is useful to understand mathematical problems that arise in the Navier-Stokes equations under certain conditions, such as limit of small viscosities, formation of shock waves, and so on. This is the simplest model that combines non-linear advection and diffusion. Eliminating the diffusion term, the equation exhibits discontinuities as the shock waves are formed. The presence of the diffusion term counteracts the effect of non-linearity, resulting in an equilibrium between the non-linear advection term and linear diffusion.
\begin{equation} \label{burger_eq}
 \dfrac{\partial{u}}{\partial{t}}+u\dfrac{\partial{u}}{\partial{x}}=\mu \dfrac{\partial^{2}{u}}{\partial^{2}{x}}
\end{equation}

\begin{figure}[H]
\centering
\includegraphics[width=0.8\linewidth]{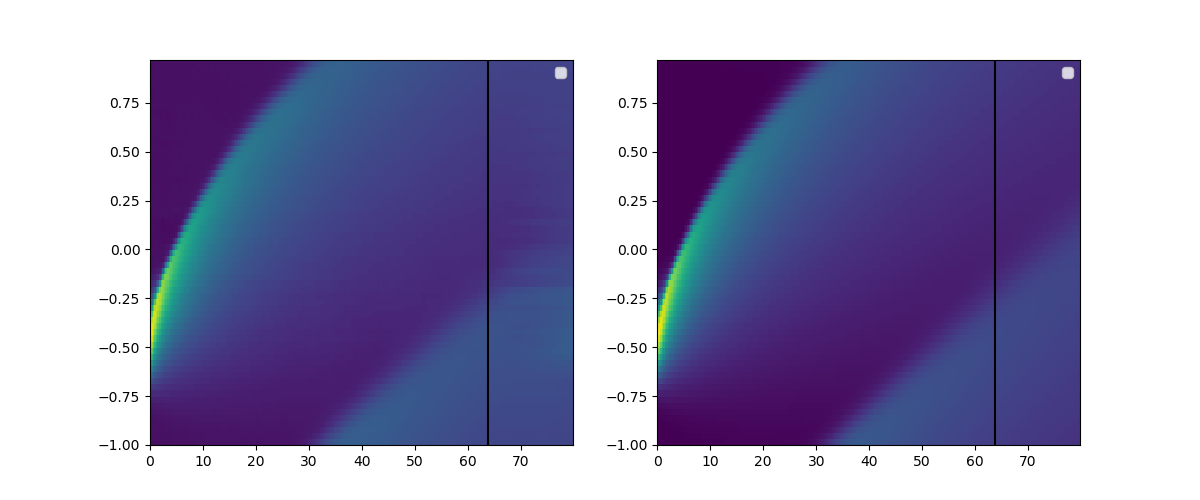}
\caption{x vs t Burgers equation dynamics. Learnt on the left and ground truth on the right.}
\label{burgers}
\end{figure}

\begin{figure}[H]
\centering
\begin{subfigure}[h]{0.48\linewidth}
\includegraphics[width=\linewidth]{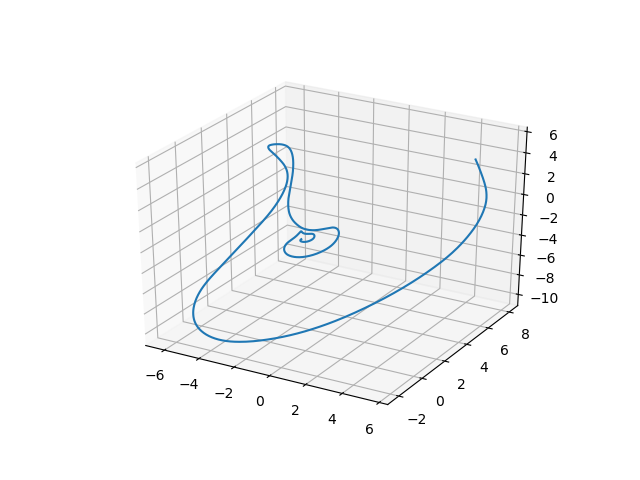}
\end{subfigure}
\begin{subfigure}[h]{0.48\linewidth}
\includegraphics[width=\linewidth]{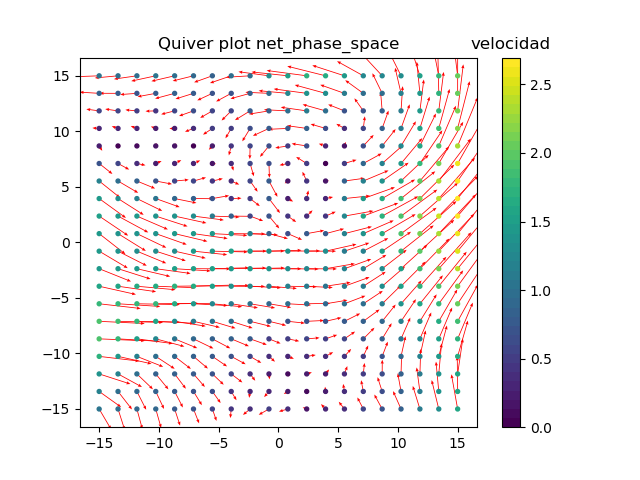}
\end{subfigure}
\caption{Latent phase space and two principal component projection of the learnt latent phase space for the training and test time together.}
\label{burgers_moving}
\end{figure}

In figure \ref{burgers}, both learnt and ground truth dynamics are shown for the Burgers equation. In figure \ref{burgers_moving}, it is clear where the formation of the shock wave happens, as well as how the dissipative dynamics of the underlying system are correctly captured.

As mentioned above, once the training phase has been completed, the integration of the trajectories has been done using an off-the-shelf ODEs integrator (\texttt{lsoda} BDF-4, specifically). The latent derivative of the trained network is passed as the system derivative function, together with the initial condition, $X_{0}$, encoded into $h_{0}$ coordinates through the encoder network.\\

\textbf{Korteweg-de-Vries}\\

Korteweg-de-Vries equation \eqref{KdV} was first conceived as model for waves on shallow water surfaces. It describes the behaviour of stable self-reinforcing wave-packets called solitons, that arise from the equilibrium of dispersive and non-linear effects in the medium. 
\begin{equation}\label{KdV}
 \dfrac{\partial{u}}{\partial{t}}+ \dfrac{\partial^{3}{u}}{\partial^{3}{x}}-u \dfrac{\partial{u}}{\partial{x}}=0
\end{equation}

This is one of the few PDEs with a known exact solution \eqref{KdV_sol}, corresponding to a functional family of soliton waves of different amplitude and speed.
\begin{equation} \label{KdV_sol}
\phi(x,t)=-\dfrac{1}{2}c\sec^{2}\left(\dfrac{\sqrt{c}}{2}(x-ct-a)\right)
\end{equation}

\begin{figure}[H]
\centering
\includegraphics[width=0.8\linewidth]{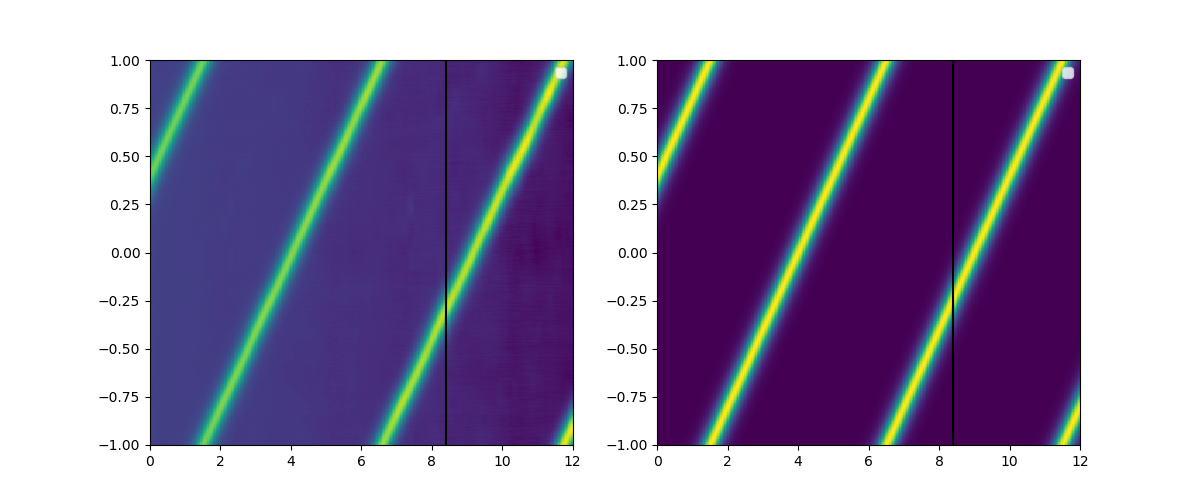}
\caption{x vs t One soliton learnt dynamics in the left side and ground truth dynamics in the right side. The black line means the training-test.}
\label{k1}
\end{figure}

In figure \ref{k1}, learnt and ground truth dynamics are represented for one soliton.

\begin{figure}[H]
\centering
\begin{subfigure}[h]{0.48\linewidth}
\includegraphics[width=\linewidth]{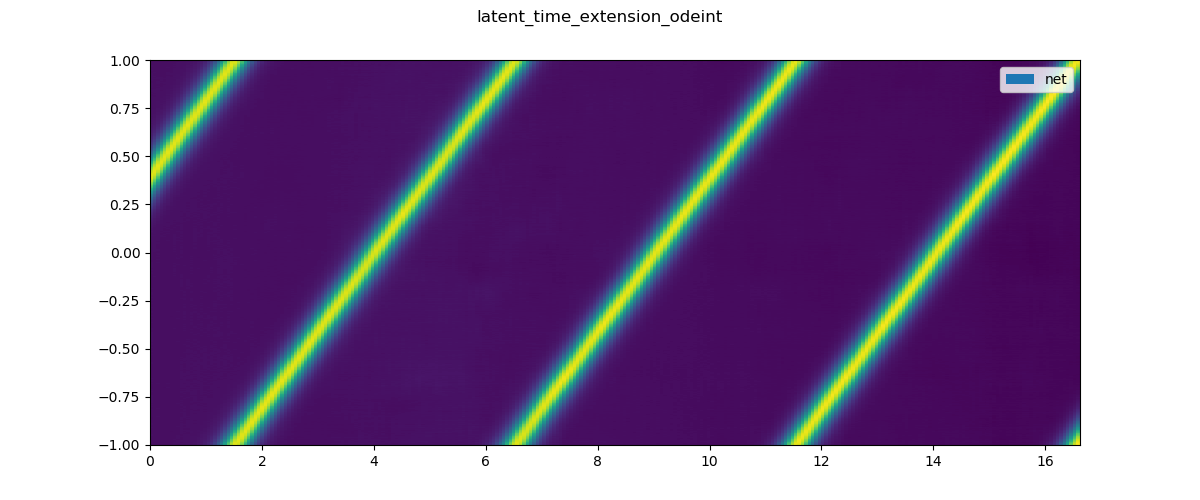}
\end{subfigure}
\begin{subfigure}[h]{0.48\linewidth}
\includegraphics[width=\linewidth]{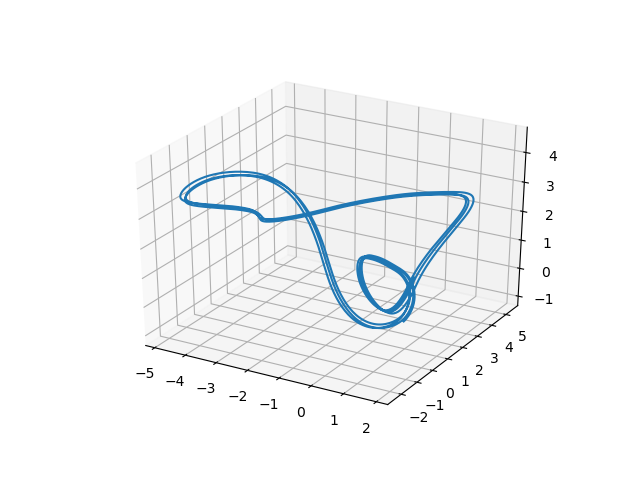}
\end{subfigure}
\caption{Time extension of the network integrated trajectory showing stable dynamics. Latent trajectory on the right side, corresponding to a conservative closed orbit, with a loop encoding the periodic boundary condition.}
\label{k2}
\end{figure}

In figure \ref{k2} the extended network integration of the learnt dynamics is shown in the $X$ and $h$ spaces, where the conservative character of the equation has been captured by the network.

\begin{figure}[H]
\centering
\includegraphics[width=0.9\linewidth]{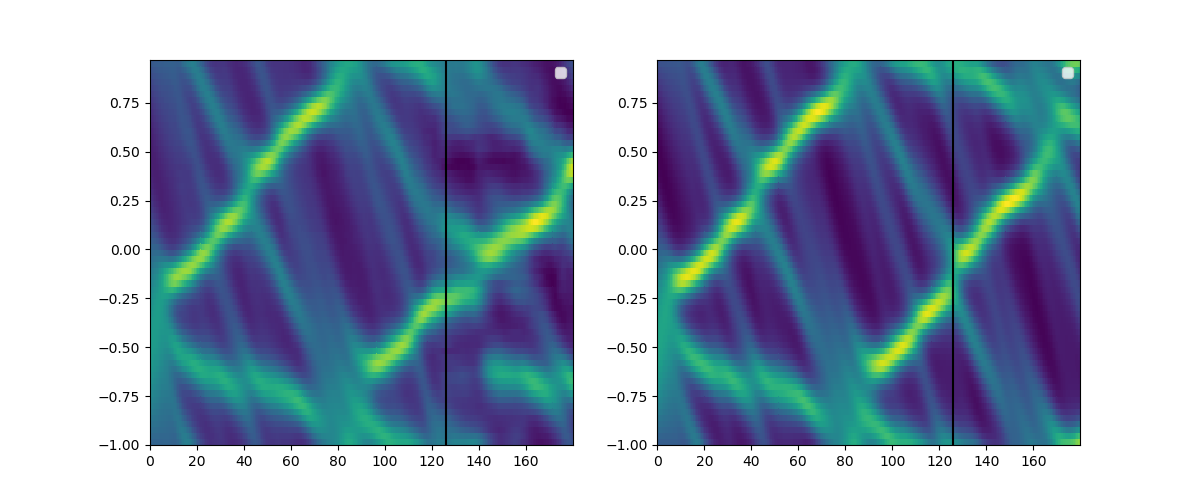}
\caption{x vs t Korteweg-de-Vries solution for an initial condition that splits mainly in three solitons. Learnt dynamics in the left side and real dynamics in the right side. Black line means the training-test.}
\label{k11}
\end{figure}

\begin{figure}[H]
\centering
\begin{subfigure}[h]{0.4\linewidth}
\includegraphics[width=\linewidth]{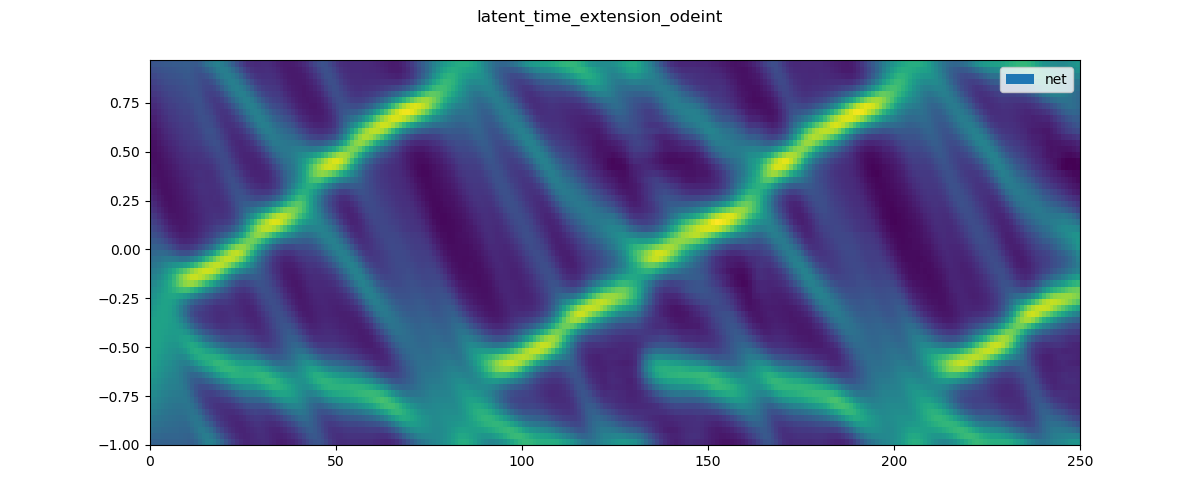}
\end{subfigure}
\begin{subfigure}[h]{0.5\linewidth}
\includegraphics[width=\linewidth]{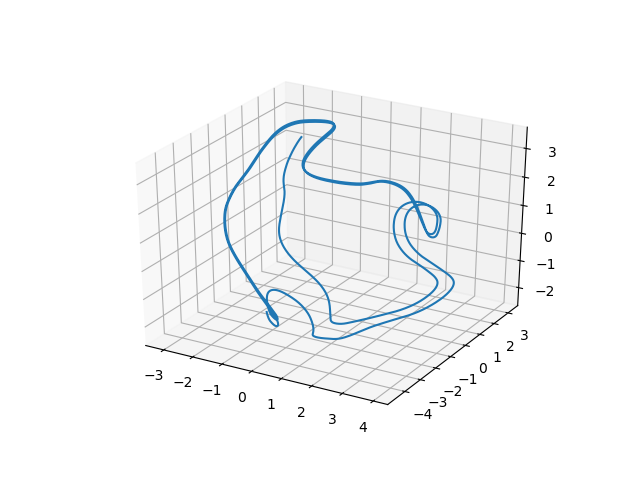}
\end{subfigure}
\caption{Network time extension of the training trajectory, showing stable conservative dynamics. After the initial gaussian transient reachs the limit cycle.}
\label{k22}
\end{figure}

As it was done for one soliton, in the case of several solitons, figures \ref{k11} and \ref{k22} are obtained. In figure \ref{k22} a two times training time latent trajectory is shown, proving as well the conservative character of the learnt solution, that appears in the latent space as a limit cycle.\\


\begin{figure}[H]
\centering
\makebox[\textwidth][c]{
\begin{subfigure}[h]{0.9\linewidth}
\makebox[\textwidth][c]{\includegraphics[width=\linewidth]{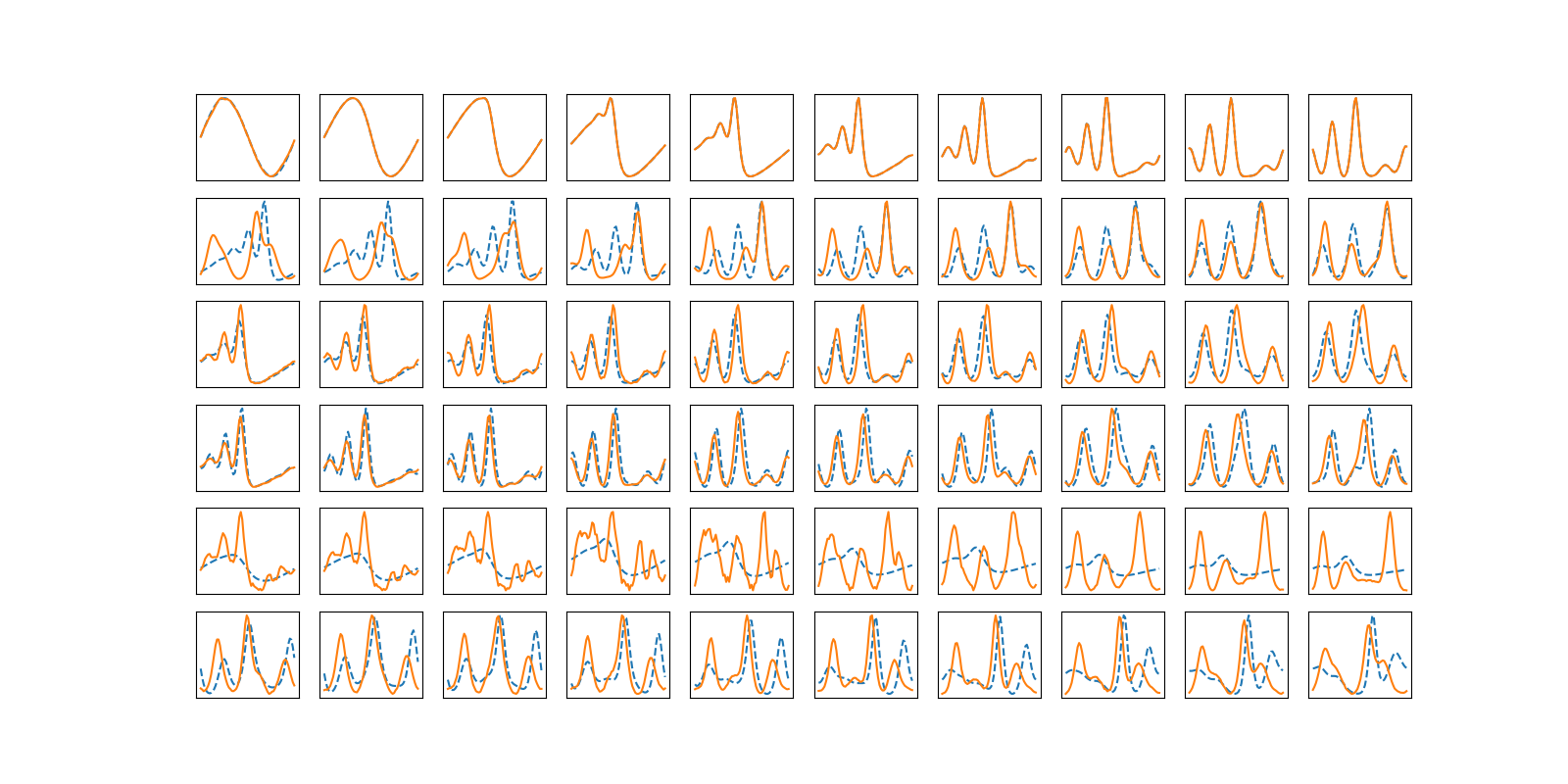}}
\end{subfigure}}
\begin{subfigure}[h]{0.45\linewidth}
\includegraphics[width=\linewidth]{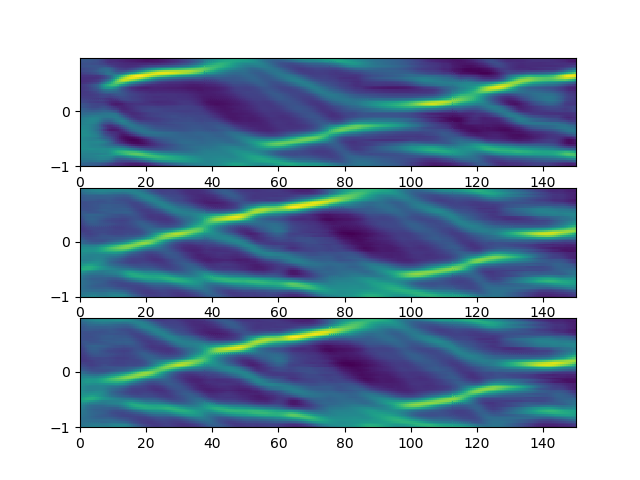}
\end{subfigure}
\begin{subfigure}[h]{0.45\linewidth}
\includegraphics[width=\linewidth]{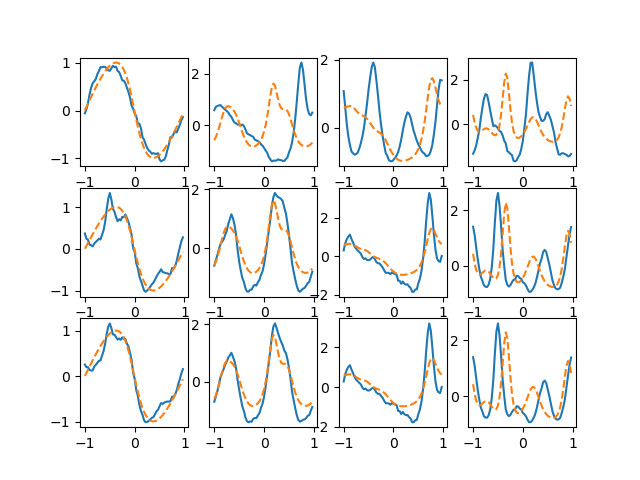}
\end{subfigure}
\caption{In the upper side are represented non-seen test solutions, the first row belongs to the training trajectory. Down it is shown the mesh plot and time snapshots of different latent generated trajectories, moving in the first principal component axis of the latent activation space. In orange the training trajectory and the generated one in blue.}
\label{k44}
\end{figure}

\textbf{Kuramoto Sivashinsky}\\

Kuramoto Sivashinsky equation \eqref{KS} is a fourth-order nonlinear PDE, that models laminar flame fronts with diffusive instabilities. The sign of the diffusive term acts as an energy source and it has a destabilising effect. The convective $uu_{x}$ term transfers energy to smaller wavelengths, where the fourth order derivative term dominates and acts as a stabilizing factor, as it is explained with the dispersion relation $-iw=k^{2}-k^{4}$. This richer set of mechanisms to transfer energy, can lead to very complex behaviours, including chaoticity. Kuramoto Sivashinsky is the simplest known PDE that exhibits this phenomena.
\begin{equation}\label{KS}
 \dfrac{\partial{u}}{\partial{t}}+ \dfrac{\partial^{4}u}{\partial^{4}x}+\dfrac{\partial^{2}{u}}{\partial^{2}{x}}+u\dfrac{\partial u}{\partial x}=0
\end{equation}

\begin{figure}[H]
\centering
\begin{subfigure}[h]{0.4\linewidth}
\includegraphics[width=\linewidth]{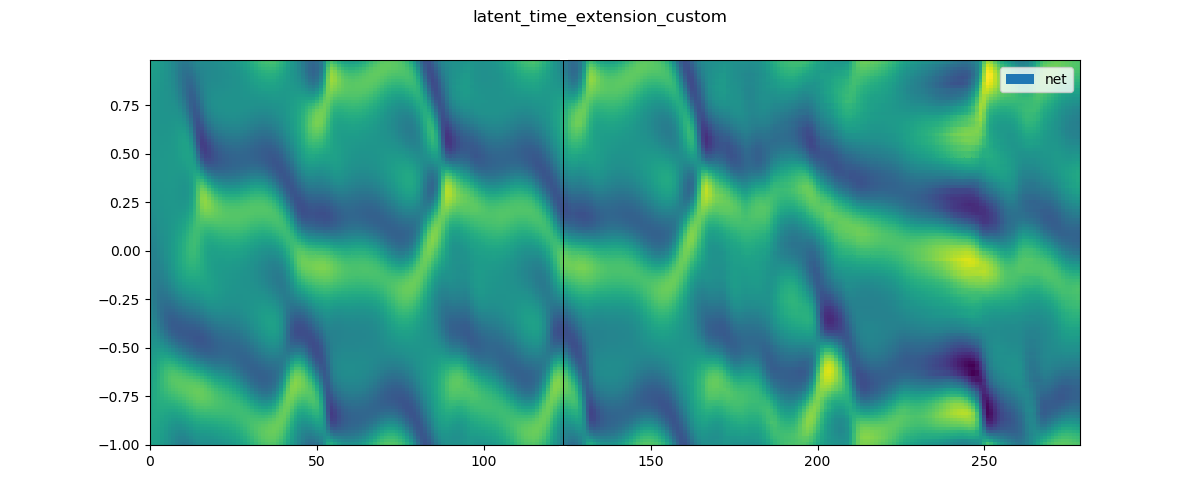}
\end{subfigure}
\begin{subfigure}[h]{0.4\linewidth}
\includegraphics[width=\linewidth]{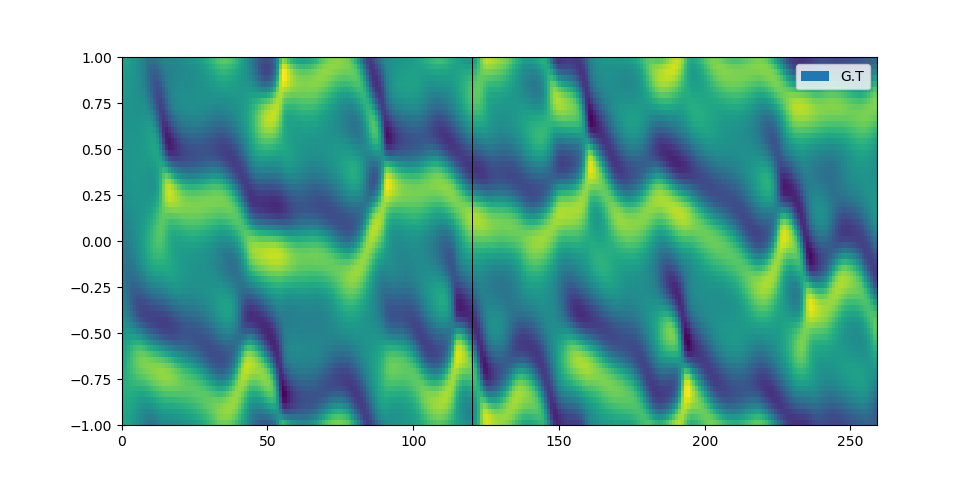}
\end{subfigure}
\caption{In the upper side,x vs t extended network dynamics. In the lower side, real dynamics. Black line means the training-test.}
\label{ks22}
\end{figure}

\begin{figure}[h]
\centering
\includegraphics[width=0.9\linewidth]{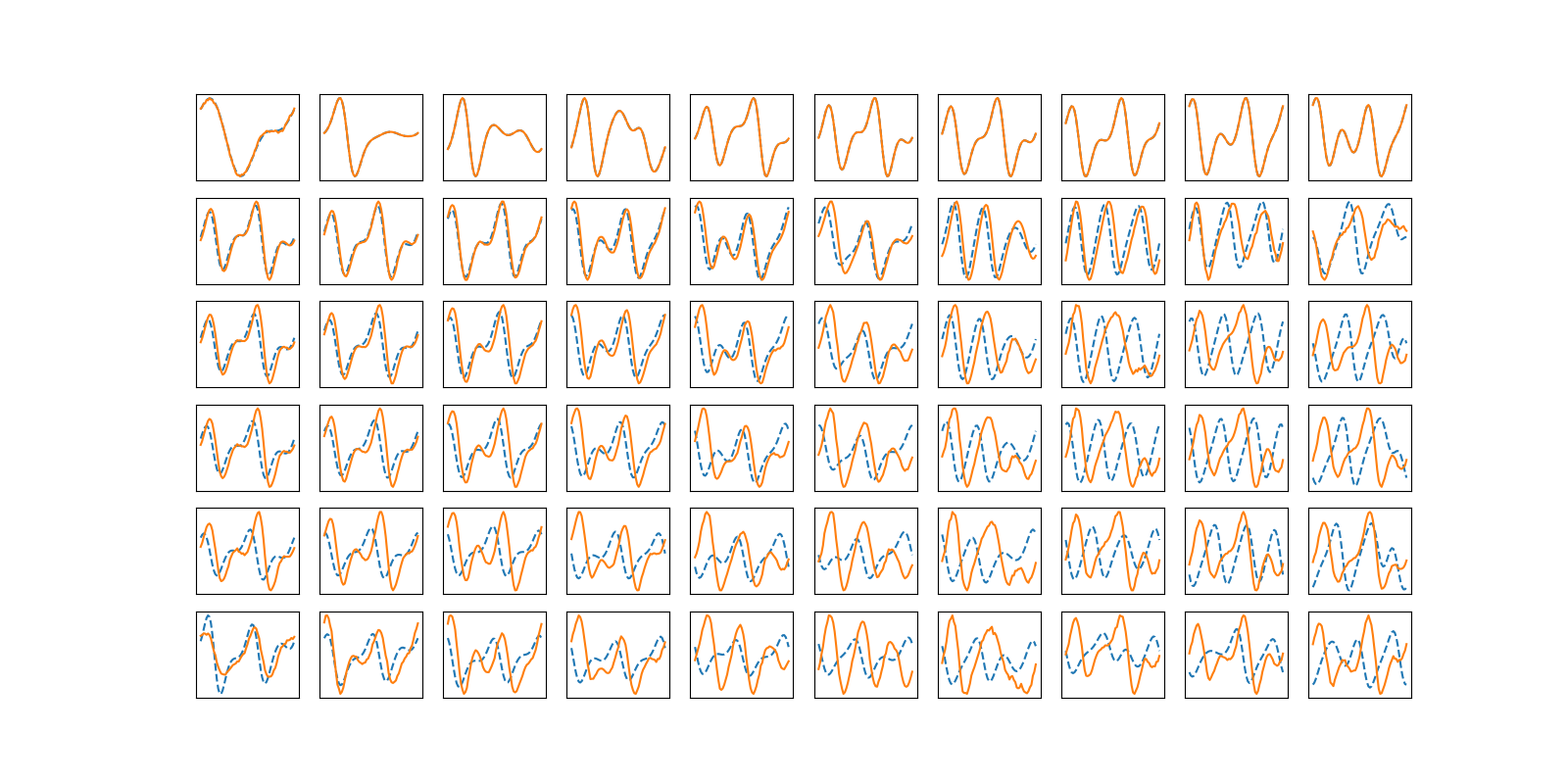}
\caption{Time snapshots of different test solutions. First row corresponds to the training one. Neural network results in orange and ground truth in blue.}
\label{ks66}
\end{figure}

By using $L=22$ as the domain width and the same initial condition solutions, Kuramoto Sivashinsky has a positive Lyapunov exponent and exhibits chaotic behaviour. The intrinsic Hausdorff dimension of the chaotic atractor is greater than $3$, so encoding it in a 3D space becomes complicated. In figure \ref{ks22} it is shown the long term prediction capability of a $4$-neuron architecture. In figure \ref{ks66} in shown as in chaotic dynamics, variations are expected for slightly different initial conditions.

%% file: sections/Conclusion.tex
\section{Conclusion}
In this work it was shown that a feed-forward properly regularized neural network, recast as a recurrent neural network and trained with the appropriate loss function, is able to qualitatively learn the global features of the phase space of dynamical systems, whose trajectories it was trained with. The examined characteristics were:
\begin{enumerate}
    \item Dissipative or conservative dynamics.
    \item Limit cycles.
    \item Unseen atractors.
    \item Frequency response.
    \item Parametric bifurcations.
\end{enumerate}

It was found that even very small neural networks with non-linear $max(0,x)$ activations can exhibit complex behaviors. This fact allows them to qualitatively fit into the mold of very different dynamics, also when the difference between the original and learnt system derivative, in terms of MSE, is considerable. This promotes the possibility of using simple networks with adjustable precision to replicate the parts of a system's dynamics that interest us, with a much lower computational cost.

With the Lorentz system experiment, it became apparent how a neural network could faithfully learn a parameterized system, with little or no added capacity with respect to the single configuration system learning problem.

Advancing in the work, we have been able to encode non-linear 1D PDEs derivative into a very compact set of hidden variables. This has enabled us to solve a system of ODEs with much fewer degrees of freedom.

It was demonstrated that the encoded space was capable of solving trajectories with a variable degree of proximity to the learned one. As well as encoding complex abstract characteristics of motion such as advection velocities, dissipation rates or transient times.

Finally, in this reduced representation, the requirements for temporal discretization, that ensure stability, were significantly relaxed compared to the spectral solver used.